\documentclass[authoryear]{elsarticle}
\pdfoutput=1

\usepackage{amsmath}
\usepackage{amsfonts}
\usepackage{epsfig}
\usepackage{graphicx}
\usepackage[utf8]{inputenc}
\usepackage{url}
\usepackage{siunitx}
\usepackage{nomencl}
\usepackage[modulo]{lineno}

\makenomenclature

\newcommand{\dd}[2]{\frac{\partial #1}{\partial #2}}
\def\bs{\boldsymbol}

\def\diff{\mathrm{d}}

\makeatletter
\def\ps@pprintTitle{%
 \let\@oddhead\@empty
 \let\@evenhead\@empty
 \def\@oddfoot{}%
 \let\@evenfoot\@oddfoot}
\makeatother

\begin{document}
\begin{frontmatter}

\title{Dry deposition model for a microscale aerosol dispersion solver based on the moment method}

\author[addr1]{Viktor Šíp\corref{cor1}}
\ead{viktor.sip@fs.cvut.cz}
\author[addr1]{Luděk Beneš}

\address[addr1]{Department of Technical Mathematics, Faculty of Mechanical Engineering, Czech Technical University in Prague. Karlovo náměstí 13, 121 35 Prague 2, Czech Republic.}
\cortext[cor1]{Corresponding author}

\begin{abstract}
A dry deposition model suitable for use in the moment method has been developed. Contributions from five main processes driving the deposition - Brownian diffusion, interception, impaction, turbulent impaction, and sedimentation - are included in the model.
The deposition model was employed in the moment method solver implemented in the OpenFOAM framework.
Applicability of the developed expression and the moment method solver was tested on two example problems of particle dispersion in the presence of a vegetation on small scales: a flow through a tree patch in 2D and a flow through a hedgerow in 3D.
Comparison with the sectional method showed that the moment method using the developed deposition model is able to reproduce the shape of the particle size distribution well. The relative difference in terms of the third moment of the distribution was below 10\% in both tested cases, and decreased away from the vegetation. Main source of this difference is a known overprediction of the impaction efficiency.
When tested on the 3D test case, the moment method achieved approximately eightfold acceleration compared to the sectional method using 41 bins.

\end{abstract}


\end{frontmatter}

\section{Introduction}

Urban vegetation is receiving a significant amount of attention from researchers in recent years. This interest stems from its impacts on the environment, affecting the pedestrian comfort and mitigating the negative health effects of the air pollution \citep{LitschkeKuttler08,Janhall15}.

Microscale modelling using the computational fluid dynamics (CFD) proved to be an indispensable tool for assessing the impacts of the vegetation in the urban settings. Some numerical studies focused only on the effects of the vegetation on the flow \citep{KenjeresKuile13} or on the thermal comfort of the pedestrian \citep{MochidaLun08}.
Others investigated pollutant dispersion in the presence of the vegetation, but without taking the deposition of the pollutant into account \citep{JeanjeanEtAl15,GromkeBlocken15a,GromkeBlocken15b}.
When including the deposition, authors opted both for simplified model with constant deposition velocity \citep{VranckxEtAl15} and complex models expressing the dependence of the deposition velocity on various parameters such as the particle size or the wind speed \citep{TiwaryEtAl05,Steffens12}.

Dispersion of the particles with a fixed size can be described by one scalar partial differential equation. When the behaviour of the particle size distribution is of interest, straightforward approach - so called \textit{sectional} approach - is to divide the size range into a number of discrete bins and then model the appropriate number of scalar PDEs, i.e. one for each bin.
Other option is to use the transport equation for the moments of the particle size distribution.
Such approach can reduce the number of PDEs to be solved, and therefore reduce the computational demands.
This class of methods, here referred to as the \textit{moment method}, has been used for the simulation of the aerosol behaviour for a long time \citep{WhitbyEtAl91}.

Usage of the moment method in the air quality models is also widespread \citep[e.g.][]{BinkowskiShankar95,PirjolaEtAl99,JungEtAl03}.
Adapting the deposition velocity models to the moment method framework is not straightforward, since the mathematical formulation of the moment method requires all terms in the equation to be in the form of the power law of the particle size. \citet{BinkowskiShankar95} simplified the problem by using the resistance model with Brownian particle diffusivity and settling velocity averaged over the particle size range. \citet{BaeEtAl09} developed a deposition velocity model based on the model proposed by \citet{Raupach01a}. This model, however, only includes the processes of Brownian diffusion, impaction and gravitational settling, and does not take into account the processes of interception and turbulent impaction, which play an important role in the dry deposition process \citep{PetroffEtAl08b}.

This study aims to fix this shortcoming by adapting the model by \citet{PetroffEtAl08b} for the use in the moment method.
The developed model is then used in the microscale CFD solver to solve the problems of a pollutant dispersion in the presence of a vegetation.
To the authors' best knowledge, the moment method has not yet been applied to the microscale urban vegetation problems.
Comparison of the obtained results with the results from the sectional model shows the applicability of such approach.

\section{Mathematical formulation}

\subsection{Number concentration equation}

The governing equation for the transport and the deposition of the aerosol particles of a diameter $d_p$ in the flow field given by the velocity $\bs{u}$ can be formulated as
\begin{equation}
  \dd{n(d_p)}{t} =
  - \underbrace{\nabla \cdot \bs{u} n(d_p)}_{\textrm{Convection}}
  + \underbrace{\nabla \cdot D \nabla n(d_p)}_{\textrm{Diffusion}}
  - \underbrace{\nabla \cdot \bs{u}_s(d_p) n(d_p)}_{\textrm{Gravitational settling}}
  - \underbrace{\mathrm{LAD} u_d(d_p) n(d_p)}_{\textrm{Deposition}},
  \label{eq:n-eq}
\end{equation}
where $n(d_p)$ is the number concentration of the particles \citep{WhitbyMcMurry97}.
Diffusion coefficient $D = \nu_T / Sc_T$ is expressed as a fraction of the turbulent viscosity and the turbulent Schmidt number.
Effects of the gravitational acceleration $\bs{g}$ are captured in the terminal settling velocity of a particle, given by the Stokes' equation,
\begin{equation}
  \bs{u}_s(d_p) = \frac{d_p^2 \rho_p \bs{g} C_C}{18 \mu},
  \label{eq:us}
\end{equation}
where $\rho_p$ is the density of the particle, $\mu$ is the dynamic viscosity of air, and $C_C$ is the Cunningham correction factor \citep{Hinds99}. The formula used for the correction factor is discussed in section \ref{sec:brown-dif}.

The removal of the particles via dry deposition is modelled by a product of three parameters: \textit{leaf area density} (LAD), defined as a leaf area per unit volume (\si{\meter^2 \meter^{-3}}), deposition velocity $u_d$ (\si{\m \per \s}) measuring the filtration efficiency of the vegetation under given conditions, and the particle concentration \citep{Raupach01}. Its form is discussed in section \ref{sec:depvel}.

\subsection{Moment equations}
\label{sec:moment-eqs}

The moment method is based on the idea that in order to model the size distribution of the particles, we can investigate the behaviour of the moments of the distribution.
Moment of the distribution is defined as
\begin{equation}
  M_k = \int_0^\infty d_p^k n(d_p) \diff d_p,
  \label{eq:def-moment}
\end{equation}
where $k$ is the order of the moment. Some moments have straightforward physical interpretation: $M_0 = \int_0^\infty n(d_p) \diff d_p = N$ is the total number concentration, $M_2 = \int_0^\infty d_p^2 n(d_p) \diff d_p = 1/\pi \cdot S$ is proportionate to the surface area concentration and $M_3 = \int_0^\infty d_p^3 n(d_p) \diff d_p = 6/\pi \cdot V$ is proportionate to the volume concentration.

Assuming $n(d_p)$ is sufficiently smooth in space and time, moment equations are obtained by multiplying Eq. (\ref{eq:n-eq}) by $d_p^k$, integrating over the whole size range and interchanging the derivatives and the integrals:
\begin{equation}
  \begin{split}
  \dd{M_k}{t} = &
  - \underbrace{\nabla \cdot \bs{u} M_k}_{\textrm{Convection}}
  + \underbrace{\nabla \cdot D \nabla M_k}_{\textrm{Diffusion}}
  - \underbrace{\int_0^\infty d_p^k \nabla \cdot \bs{u}_s(d_p) n(d_p) \diff d_p}_{\textrm{Gravitational settling}} \\
 & - \underbrace{\mathrm{LAD} \int_0^\infty d_p^k u_d(d_p) n(d_p) \diff d_p}_{\textrm{Deposition}}.
  \end{split}
  \label{eq:moment-eq}
\end{equation}

Now we are left with the evaluation of the integrals in (\ref{eq:moment-eq}). This can be done easily if the multiplicative terms are in a form of a polynomial function of $d_p$. Such is the case with the gravitational term, if we take into account that gravity plays significant role only for larger particles, where the Cunningham correction factor $C_C$ in (\ref{eq:us}) can be left out. Using (\ref{eq:us}) in the third term on the RHS of (\ref{eq:moment-eq}), the term can be rewritten as
\begin{equation}
   - \nabla \cdot \bs{g} \frac{\rho_p}{18 \mu} \int_0^\infty d_p^{k+2} n(d_p) \diff d_p
   = - \nabla \cdot \bs{g} \frac{\rho_p}{18 \mu}  M_{k+2}.
  \label{eq:mom-grav}
\end{equation}
Here we introduced a dependence on the moment of a higher order. That necessitates that we either solve a separate moment equation also for this higher order moment, or that this moment can be calculated from the moments that we solve for.

The task of integrating the deposition term is more difficult and will be examined in the following section.

\subsection{Deposition model for the moment method}
\label{sec:depvel}

Variety of models describing the rate of particle transport from the air to the vegetation surface has been proposed \citep{PetroffEtAl08a}.
Dry deposition schemes used in the large-scale air quality models are usually formulated in terms of a friction velocity or a wind speed at some height above the canopy, and do not explicitely model the behaviour inside the canopy.
As such, they are not directly applicable in the microscale CFD models, but the same principles governing the deposition still apply.

\begin{figure}[h]
  \centering
  \includegraphics[width=0.5\textwidth]{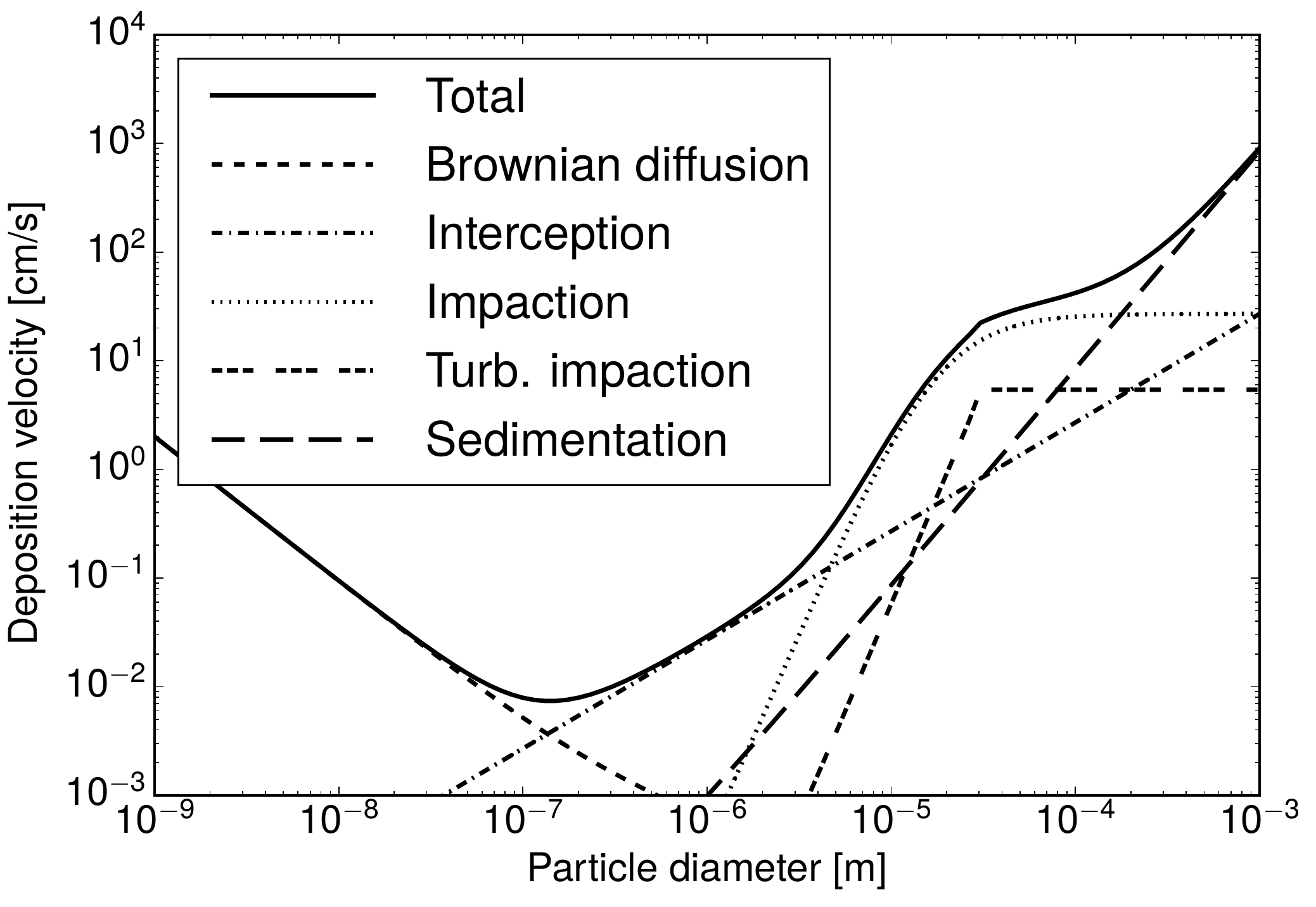}
  \caption{Dependence of the deposition velocity on the particle diameter in the original model ($\rho_p = \SI{1300}{\kilogram\ \meter^{-3}}, d_e = \SI{2}{\milli \meter}, U = \SI{1}{\m\per\s}, u_f = \SI{0.3}{\m\per\s}$. See sections \ref{sec:brown-dif} - \ref{sec:sedim} for the description of the parameters).}
  \label{fig:depvel-petroff-contribs}
\end{figure}

In this study we adopted the deposition velocity expression given by \citet{PetroffEtAl08b} for needle-like obstacles.
\citet{PetroffEtAl08b} formulated the expressions for deposition velocities (or collection velocities in the original terminology) associated with each of the five main processes driving the dry deposition inside the canopy.
The model then assumes that these  processes, which are the Brownian diffusion, interception, impaction, turbulent impaction, and sedimentation, are acting in parallel and independently.
The deposition velocity thus can be written as a sum of the deposition velocities of all processes,
\begin{equation}
  u_d = u_{BD} + u_{IN} + u_{IM} + u_{TI} + u_{SE}.
\end{equation}
Contributions of each process to the deposition velocity for an exemplary set of parameters are shown on Fig. \ref{fig:depvel-petroff-contribs}.

The assumption of the parallel and independent acting is advantageous for adapting the model to the moment method, since it allows us to split the rightmost integral in Eq. (\ref{eq:moment-eq}) into integrals pertaining to the every physical process separately.
In this section, we will describe each process in more detail and show how it can be adapted to the moment method.

Few simplifications were made to the original model to make the subsequent analysis simpler: we considered only plagiophile canopies and Dirac distribution of the needle sizes.
Also note that here we focus only on needle-like obstacles. Similar model for broadleaf canopies, given in \citep{PetroffEtAl09}, could be adapted to the moment method as well.

\subsubsection{Brownian diffusion}
\label{sec:brown-dif}

Brownian diffusion is the dominant process driving the deposition of the particles smaller then 0.1 \si{\micro \meter} \citep{LitschkeKuttler08}.
Original model formulates the contribution to the deposition velocity due to the Brownian diffusion as
\begin{equation}
  u_{BD} = U C_B Sc^{-2/3} Re^{n_B - 1}
  \label{eq:vd-B}
\end{equation}
where $U$ is the magnitude of the wind velocity, $Sc = \nu_a/D_B$ is the Schmidt number (with $\nu_a$ being the kinematic viscosity of air and $D_B$ the Brownian diffusion coefficient, $D_B = (C_C k_b T_a)/(3 \pi \mu_a d_p)$), $Re = U d_e/\nu_a$ is the Reynolds number, and $d_e$ is the needle diameter.
Assuming laminar boundary layer around the obstacles, the model constants hold the values $n_B = 0.5$ and $C_B = 0.467$.

\begin{figure}[h]
  \centering
  \includegraphics[width=0.8\textwidth]{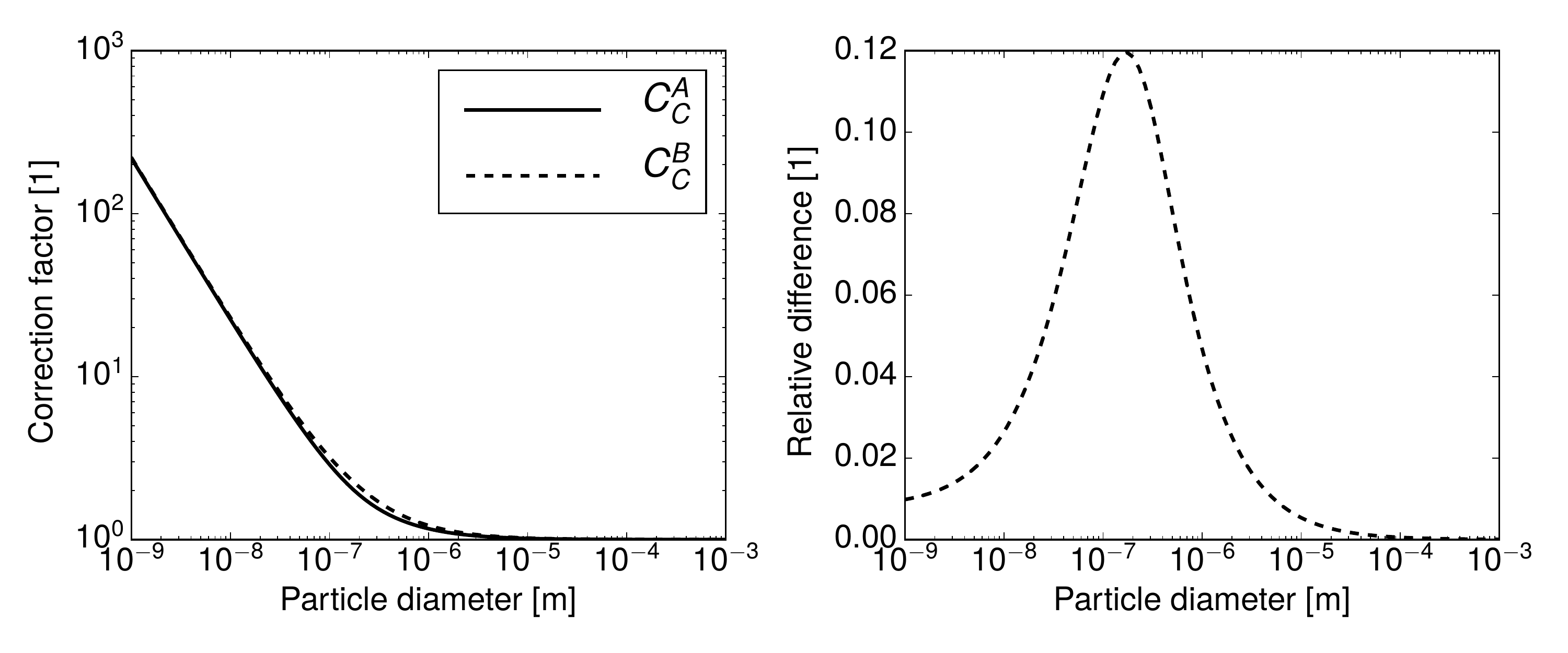}
  \caption{(Left) Two expressions for the Cunningham correction factor. (Right) Relative difference $(C_C^B - C_C^A)/C_C^A$.}
  \label{fig:correction-factor}
\end{figure}

Cunningham correction factor $C_C^A = 1 + 2\lambda/d_p  (1.257 + 0.4 \exp(-1.1 d_p/2\lambda))$ is used in the original model \citep{PetroffEtAl08b}, where $\lambda$ is the mean free path of the particle in the air.
In this whole study we use simpler approximation $C_C^B = 1 + 3.34 \lambda/d_p$ \citep{BaeEtAl09}.
Comparison of these expressions is on Fig. \ref{fig:correction-factor}, where it can be seen that their difference peaks to 12\% for particle diameter around 0.2 \si{\micro\meter}.

Furthermore, for the Brownian diffusion we take into account only the size-dependent part of the correction factor, dominant in the particle size range where the diffusion is significant, $C_C \approx 3.34 \frac{\lambda}{d_p}$. Putting the expressions above into Eq. (\ref{eq:vd-B}), we obtain
\begin{equation}
  u_{BD} \approx u_{BD}' = U^{n_B} \gamma_1 \gamma_2^{2/3} d_p^{-4/3},
  \label{eq:vd-B-apprx}
\end{equation}
where ${\gamma_1 = C_B \left(d_e/\nu_a\right)^{n_B - 1} \left(3 \pi \nu_a^2 \rho_a/(k_b T_a)\right)^{-2/3}}$ and $\gamma_2 = 3.34 \lambda$.

Using this formula, the contribution to the moment equation can be written as
\begin{equation}
  \left. \dd{M_k}{t} \right|_{BD} = - \mathrm{LAD} \ U^{n_B} \gamma_1 \gamma_2^{2/3} M_{k - 4/3}.
  \label{eq:mom-B}
\end{equation}

\subsubsection{Interception}

Interception denotes the process where the particle follows the streamline, but too close to the obstacle so that it is captured on the surface. The original model parameterizes its contribution to the deposition velocity as
\begin{equation}
  u_{IN} = 2 U k_x \frac{d_p}{d_e},
  \label{eq:vd-IN}
\end{equation}
where $k_x = 0.27$ is the ratio of the leaf surface projected on the plane perpendicular to the flow to the total leaf surface.

Since there is a linear dependence on the particle diameter, the expression can be integrated as is, resulting in
\begin{equation}
  \left. \dd{M_k}{t} \right|_{IN} = - \mathrm{LAD} \gamma_3 U M_{k+1},
  \label{eq:mom-IN}
\end{equation}
with $\gamma_3 = 2 k_x/d_e$.

\subsubsection{Inertial impaction}
\label{sec:vd-IM}

\begin{figure}[h]
  \centering
  \includegraphics[width=0.5\textwidth]{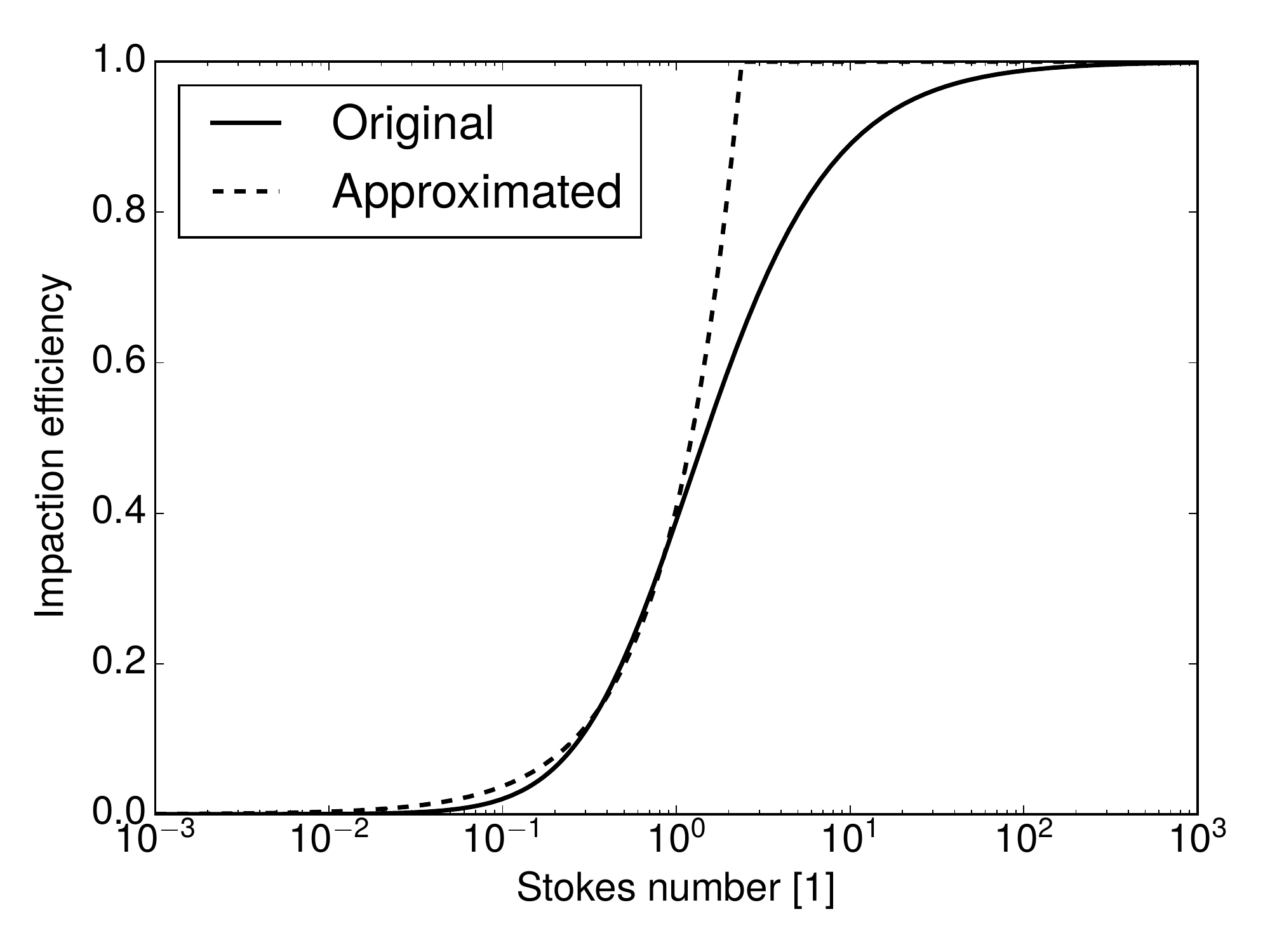}
  \caption{Original and approximated expressions for the impaction efficiency.}
  \label{fig:impaction-efficiency}
\end{figure}

Inertial impaction occurs when particles do not follow the streamlines due to their inertia, resulting in the collision with the obstacle.
The deposition velocity due to the inertial impaction is written as
\begin{equation}
  u_{IM} = U k_x E_{IM},
  \label{eq:vd-IM}
\end{equation}
where $E_{IM} = \left( \frac{St}{St + \beta} \right)^2$ is the impaction efficiency with the constant $\beta = 0.6$.
To use this deposition velocity in the moment equations, we approximate the impaction efficiency as $E_{IM}' = \min (a \ St^b, 1)$, where the coefficients $a = 0.407$ and $b = 1.039$ were obtained by minimizing the quadratic error of the original function and the approximation for $St \in [0, 1]$. The original function and the approximation are shown on Fig. \ref{fig:impaction-efficiency}.
It is clear that this approximation introduces overprediction of the impaction efficiency for Stokes numbers between 1 and 10. This was however deemed to be an acceptable compromise between the accuracy and the computational efficiency. Without the approximation, one could use the numerical integration to evaluate the integral to obtain more precise approximation of the original formula, but at a significantly higher computational cost.

Approximated deposition velocity can then be formally written as
\begin{equation}
  u'_{IM} = U k_x E_{IM}'.
  \label{eq:vd-IM-apprx}
\end{equation}
After some algebraic manipulations the corresponding term in the moment equation can be written using the incomplete moments
  $M_k^- (x) = \int_0^x d_p^k n(d_p) \diff d_p$ and $M_k^+ (x) = \int_x^\infty d_p^k n(d_p) \diff d_p$ as
\begin{equation}
  \left. \dd{M_k}{t} \right|_{IM} = - \mathrm{LAD} \ U k_x \left(
    U^b \gamma_4 M_{k+2b}^{-}(d_p^{T1}) + M_k^{+}(d_p^{T1})
    \right)
    \label{eq:mom-IM}
\end{equation}
with the threshold $d_p^{T1} = \sqrt{\frac{18 \mu_a d_e}{\rho_p U a^{1/b}}}$ and $\gamma_4 = a \left( \frac{\rho_p}{18 \mu_a d_e} \right)^b$.

\subsubsection{Turbulent impaction}

Effect of the particle impaction due to the canopy turbulence is described by the deposition velocity
\begin{equation}
u_{TI} =
\left\{
  \begin{array}{ll}
      u_f K_{TI1} {\tau_p^+}^2  & \mbox{if } \tau_p^+ < 20, \\
      u_f K_{TI2}               & \mbox{if } \tau_p^+ \geq 20.
  \end{array}
\right.
\label{eq:vd-TI}
\end{equation}
Here $\tau_p^+ = \tau_p u_f^2/\nu_a$ is the dimensionless particle relaxation time, $\tau_p = \frac{\rho_p C_c d_p^2}{18 \mu_a}$ is the particle relaxation time, $u_f$ is the local friction velocity, $K_{TI1} = 3.5 \cdot 10^{-4}$, and $K_{TI2} = 0.18$.
The contribution to the moment equation can be again expressed using the incomplete moments,
\begin{equation}
\left. \dd{M_k}{t} \right|_{TI} = -\mathrm{LAD} \left( u_f^5 \gamma_5 M_{k+4}^{-} (d_p^{T2}) + u_f K_{TI2} M_k^+ (d_p^{T2}) \right)
  \label{eq:mom-TI}
\end{equation}
with the threshold $d_p^{T2} = \sqrt{\frac{360 \mu_a \nu_a}{\rho_p u_f^2}}$ and $\gamma_5 = \frac{K_{TI1} \rho_p^2 \rho_a^2}{(18 \mu_a^2)^2}$.

\subsubsection{Sedimentation}
\label{sec:sedim}

Sedimentation plays the major role for the particles with the diameter above \SI{10}{\micro \meter}.
The sedimentation contribution to the deposition velocity is expressed as
\begin{equation}
  u_{SE} = \frac{k_z g \rho_p C_C d_p^2}{18 \mu_a},
  \label{eq:vd-SE}
\end{equation}
where $k_z = 0.22$ is the ratio of the leaf surface projected to the horizontal plane to the total leaf surface.

Substituting this expression to the integral in the moment equation, after some algebraic manipulations we obtain
\begin{equation}
  \left. \dd{M_k}{t} \right|_{SE} = -\mathrm{LAD} \gamma_6 (M_{k+2} + \gamma_2 M_{k+1})
  \label{eq:mom-SE}
\end{equation}
with $\gamma_6 = \frac{k_z g \rho_p}{18 \mu_a}$ and $\gamma_2 = 3.34 \lambda$ as before.

\subsubsection{Comparison with the original model}

The developed model can be formally written as
\begin{equation}
u_d = u_B' + u_{IN} + u_{IM}' + u_{TI} + u_{SE}
\end{equation}
using the Equations (\ref{eq:vd-B-apprx}), (\ref{eq:vd-IN}), (\ref{eq:vd-IM-apprx}), (\ref{eq:vd-TI}) and (\ref{eq:vd-SE}). In the moment method solver it is however implemented via the Equations (\ref{eq:mom-B}), (\ref{eq:mom-IN}), (\ref{eq:mom-IM}), (\ref{eq:mom-TI}) and (\ref{eq:mom-SE}).
Two major approximation were made: first, Cunningham approximation factor is replaced by its size-dependent part in the Brownian diffusion term. Secondly, an inexact power law formula is used for the impaction efficiency.

\begin{figure}[h]
  \centering
  \includegraphics[width=0.5\textwidth]{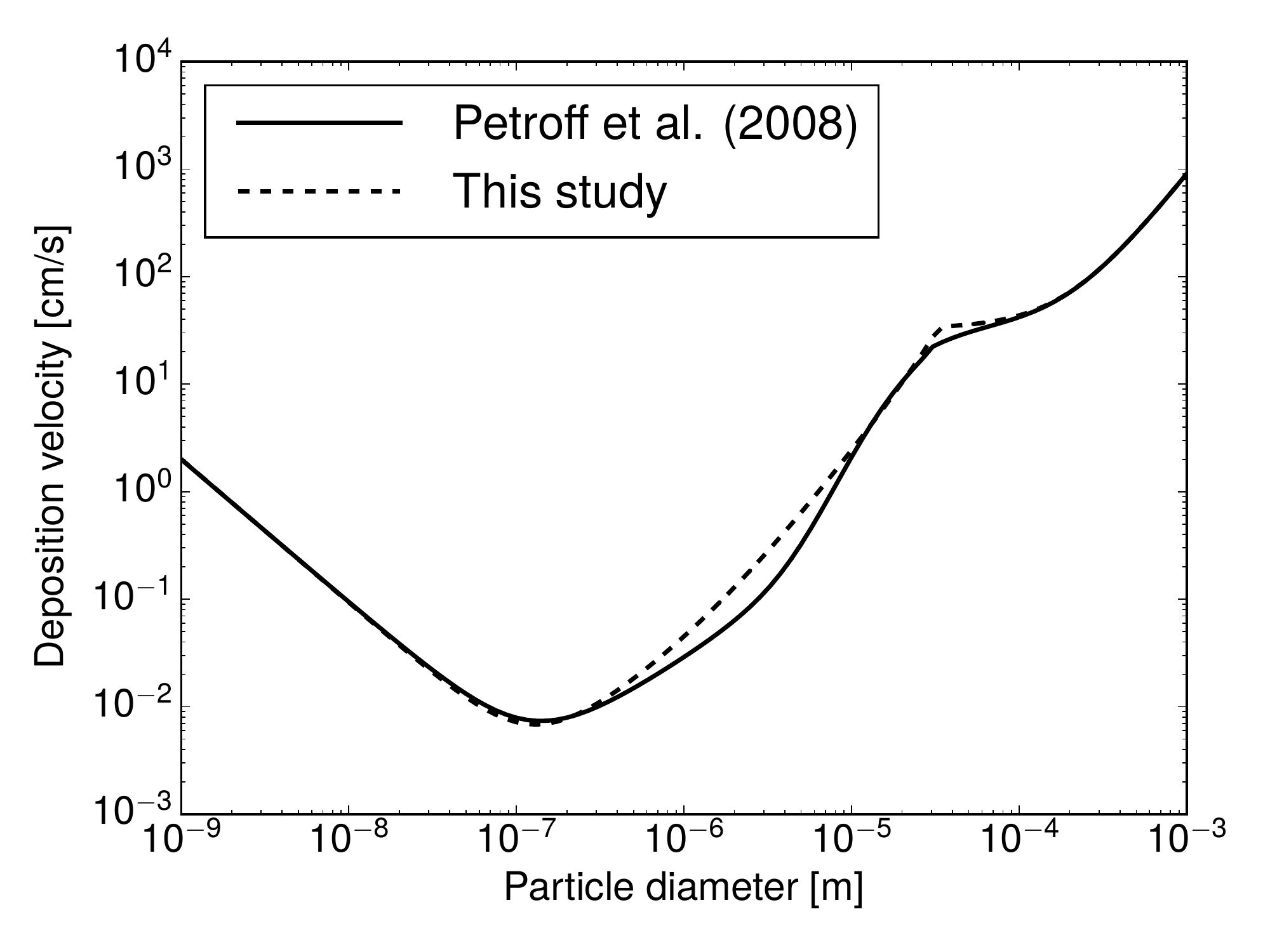}
  \caption{Comparison of the deposition velocities given by the original model and the one developed in this study ($\rho_p = \SI{1300}{\kilogram\ \meter^{-3}}, d_e = \SI{2}{\milli \meter}, U = \SI{1}{\m\per\s}, u_f = \SI{0.3}{\m\per\s}$).}
  \label{fig:depvel-comparison}
\end{figure}

Comparison of this model with the original for an exemplary set of parameters is shown on Fig. \ref{fig:depvel-comparison}.
The higher values of the deposition velocity for particles around \SI{3}{\um} are the consequence of the inexact approximation to the inertial impaction term described in section \ref{sec:vd-IM}.

The maximal difference of the deposition velocity given by the two models was determined by evaluating the deposition velocity for every combination of the parameters in the ranges expected in real-world situations ($\rho_p \in [500; 3000]\ \si{\kg\per\m\cubed}$, $d_e \in [0.5; 5]\ \si{\mm}$, $U \in [0, 10]\ \si{\m\per\s}$, $d_p \in [10^{-3}, 10^{2}]\ \si{\um}$). Each interval was discretized using 50 points. Local friction velocity $u_f$ was set to 0 $\si{\m\per\s}$, as the turbulent impaction is implemented exactly and its contribution can only reduce the relative difference of the deposition velocities.
The largest relative difference $|u_d^{orig} - u_d^{approx}|/ \mathrm{min} (u_d^{orig}, u_d^{approx})$ was found to be 4.56 for the parameter values at the end of the expected ranges ($\rho_p  = 3000\ \si{\kg\per\m\cubed}$, $d_e = 5\ \si{\mm}$, $U = 10\ \si{\m\per\s}$) and particle diameter $d_p = 5.67\ \si{\um}$, giving the deposition velocities $u_d^{orig} = 0.106\ \si{\cm\per\s}$ and $u_d^{approx} = 0.591\ \si{\cm\per\s}$.

While this difference is certainly significant, measured values shows even higher variability \citep{LitschkeKuttler08}. Considering this, the costs of the alternative to this approximation - i.e. the numerical integration of the impaction term integral - does not outweigh the better fit to the original model.

\subsection{Lognormal distribution}

Before we move on to the description of the implementation, it is necessary to provide some assumptions on the particle size distribution.
Size distributions of the atmospheric aerosols are often well fitted by a multimodal lognormal distribution \citep{SeinfeldPandis06}.
This is the distribution we will use from now on. We restrict ourselves only to the case of unimodal distribution, noting that the multimodal distribution can be modelled by a superposition of several unimodal distributions.

Unimodal lognormal distribution can be described by three parameters: total number concentration $N$, geometric mean size $d_{gn}$ and geometric standard deviation $\sigma_g$. Its probability density function is
\begin{equation}
  n(\ln d_p) = \frac{N}{\sqrt{2 \pi} \ln \sigma_g}
  \exp \left( - \frac{(\ln d_p - \ln d_{gn})^2}{2 \ln^2 \sigma_g} \right).
  \label{eq:lognormal-prob-density}
\end{equation}
Knowing the three parameters, $k$-th moment can be calculated using the formula
\begin{equation}
  M_k = N d_{gn}^k \exp \left( \frac{k^2}{2} \ln^2 \sigma_g \right).
\end{equation}
From the three moments of order 0, $k_1$ and $k_2$ the three parameters can be obtained using the relations
\begin{align}
  N & = M_0, \label{eq:n}\\
  d_{gn} & = \overline{M}_{k_1}^{\frac{1}{r (k_2 - k_1)}} \overline{M}_{k_2}^{\frac{r}{k_1 - k_2}}, \label{eq:dgn} \\
  \ln^2 \sigma_g & = \frac{2}{k_1 (k_1 - k_2)} \ln \left( \frac{\overline{M}_{k_1}}{\overline{M}_{k_2}^r} \right) \label{eq:ln2sigmag},
\end{align}
where $\overline{M}_k = \frac{M_k}{M_0}$ and $r = \frac{k_1}{k_2}$ \citep{WhitbyMcMurry97}.

For the incomplete higher order moments following holds:
\begin{align}
  M_k^- (x) = & \int_0^x d_p^k n(d_p) \diff d_p = M_k \Phi \left( \frac{\ln x - \ln d_{gn} - k \ln^2 \sigma_g}{\ln \sigma_g} \right), \label{eq:lnd-incomplete-moms-n} \\
  M_k^+ (x) = & \int_x^\infty d_p^k n(d_p) \diff d_p = M_k \left(1 - \Phi \left( \frac{\ln x - \ln d_{gn} - k \ln^2 \sigma_g}{\ln \sigma_g} \right) \right), \label{eq:lnd-incomplete-moms-p}
\end{align}
where $\Phi$ is the normal cumulative distribution function.

\subsection{Choice of the moments}

Now we turn our attention to the choice of the moments. For which orders we decide to solve the moment equation (\ref{eq:moment-eq}) is to a degree an arbitrary decision. When this problem is discussed in literature, cited reasons for a certain choice include the mathematical simplicity and ease of the formulation of the modelled processes or the physical interpretation of some moments \citep{WhitbyMcMurry97,BinkowskiShankar95}.
Choices of the moments used in the field of atmospheric aerosol modelling in the selected literature are summarized in Tab. \ref{tab:moments-choice}.

\begin{table}[h]
  \centering
  \begin{tabular}{lcp{0.4\textwidth}}
    Reference & Moments \\ \hline
    \citep{BinkowskiShankar95}   & 0, 3, 6  \\
    \citep{PirjolaEtAl99}        & 0, 2, 3  \\
    \citep{JungEtAl03}           & 0, 2, 3  \\
    \citep{KoziolLeighton07}     & 0, 1, 2  \\
    \citep{BaeEtAl09}            & 0, 3, 6  \\
  \end{tabular}
  \caption{Choices of the moments in the selected literature}
  \label{tab:moments-choice}
\end{table}
The recurrent usage of zeroth order moment brings substantial advantage, as it is equal to the total number concentration, and it is the order we will use as well. On the choice of the other moments authors differ.

To assess the influence of the choice of the moments, following numerical experiment was performed. We investigated the particle deposition in a 1D tube, spanning between 0 and 300 m. Homogenous vegetation block of LAD~=~3~\si{\meter^2 \meter^{-3}} was placed between 100 a 150 m. Velocity of the air in the whole tube was set to constant 1~\si{\m\per\s}, unaffected by the vegetation. Source of the pollutant was placed at 50 m from the inlet with the intensity of number of particles 1 \si{\s^{-1}} and the distribution parameters $\sigma_g = 0.7, d_{gn} = 3 \si{\um}$. The tube was discretized using 400 cells.

Beside the choices mentioned in Tab. \ref{tab:moments-choice}, we tested also a variant with a negative order moment: 0, -1, 1. Non integer choices of the orders would also be possible to use, but we saw no advantage that such choice could bring.

Transport and the deposition of the pollutant was calculated by the sectional model based on the Eq. (\ref{eq:n-eq}) and by the moment method based on the Eq. (\ref{eq:moment-eq}) (see section \ref{sec:implementation} for details on the implementation). To discard possible errors due to the inexact approximation of the deposition velocity, only the sedimentation contribution, adapted exactly, was taken into account.
The numerical experiment is not meant to model any real-world situation, rather just demonstrate the behaviour of the moment method in a simple setting.

\begin{figure}[h]
  \centering
  \includegraphics[width=0.48\textwidth]{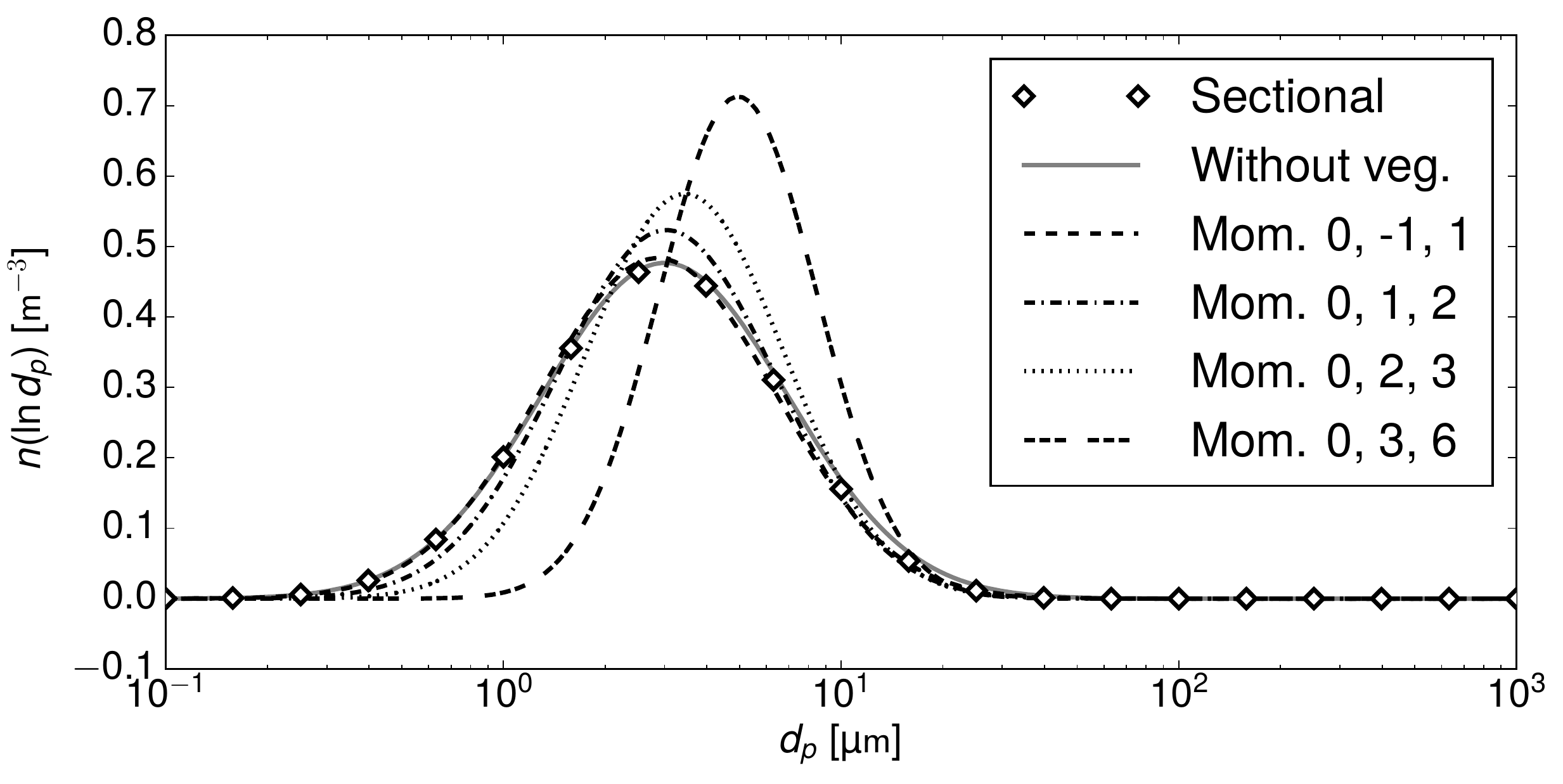}
  \includegraphics[width=0.48\textwidth]{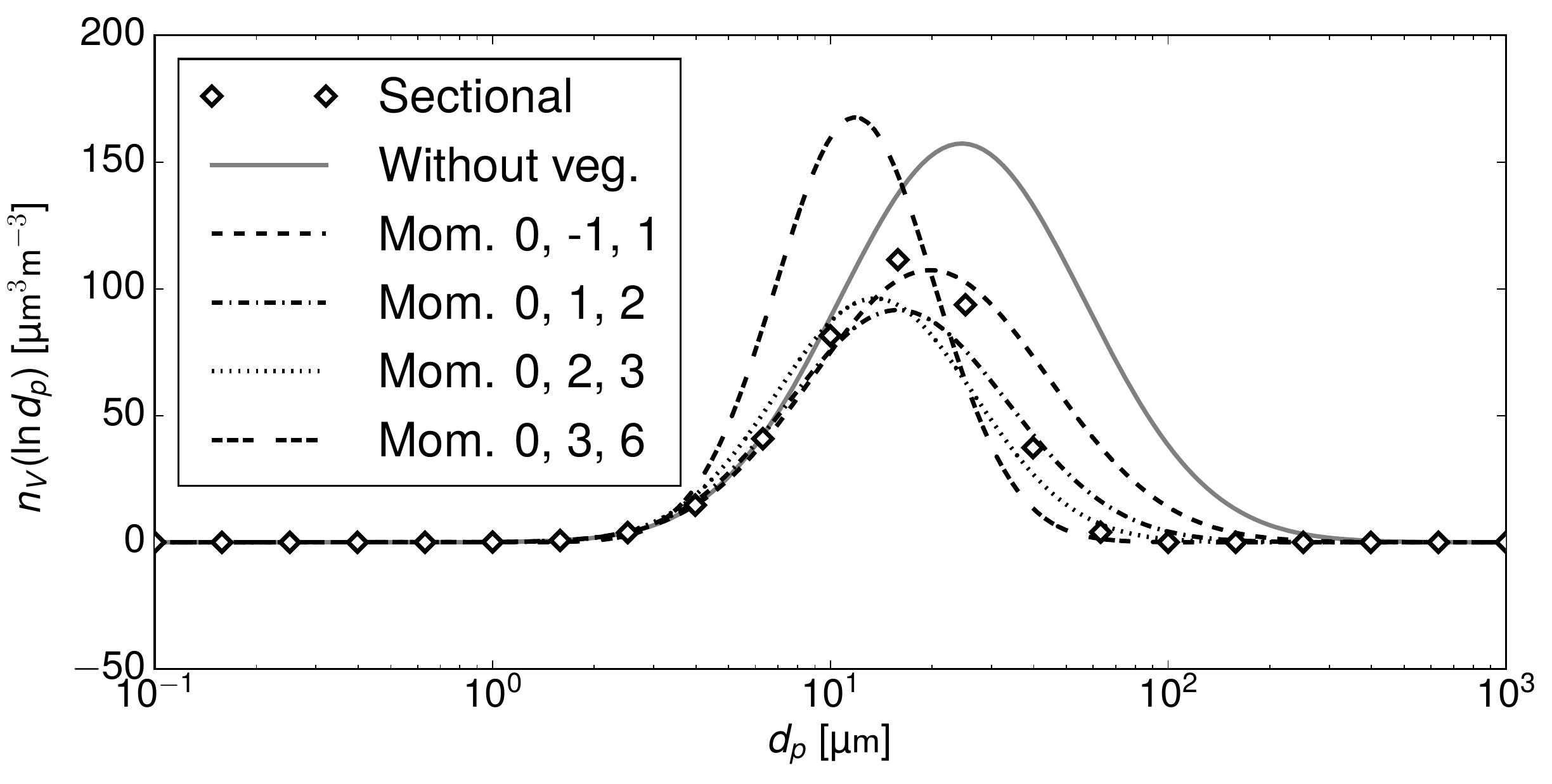}
  \caption{(Left) Number concentration (Right) Volume concentration}
  \label{fig:choice-dist}
\end{figure}

\begin{figure}[h]
  \centering
  \includegraphics[width=0.48\textwidth]{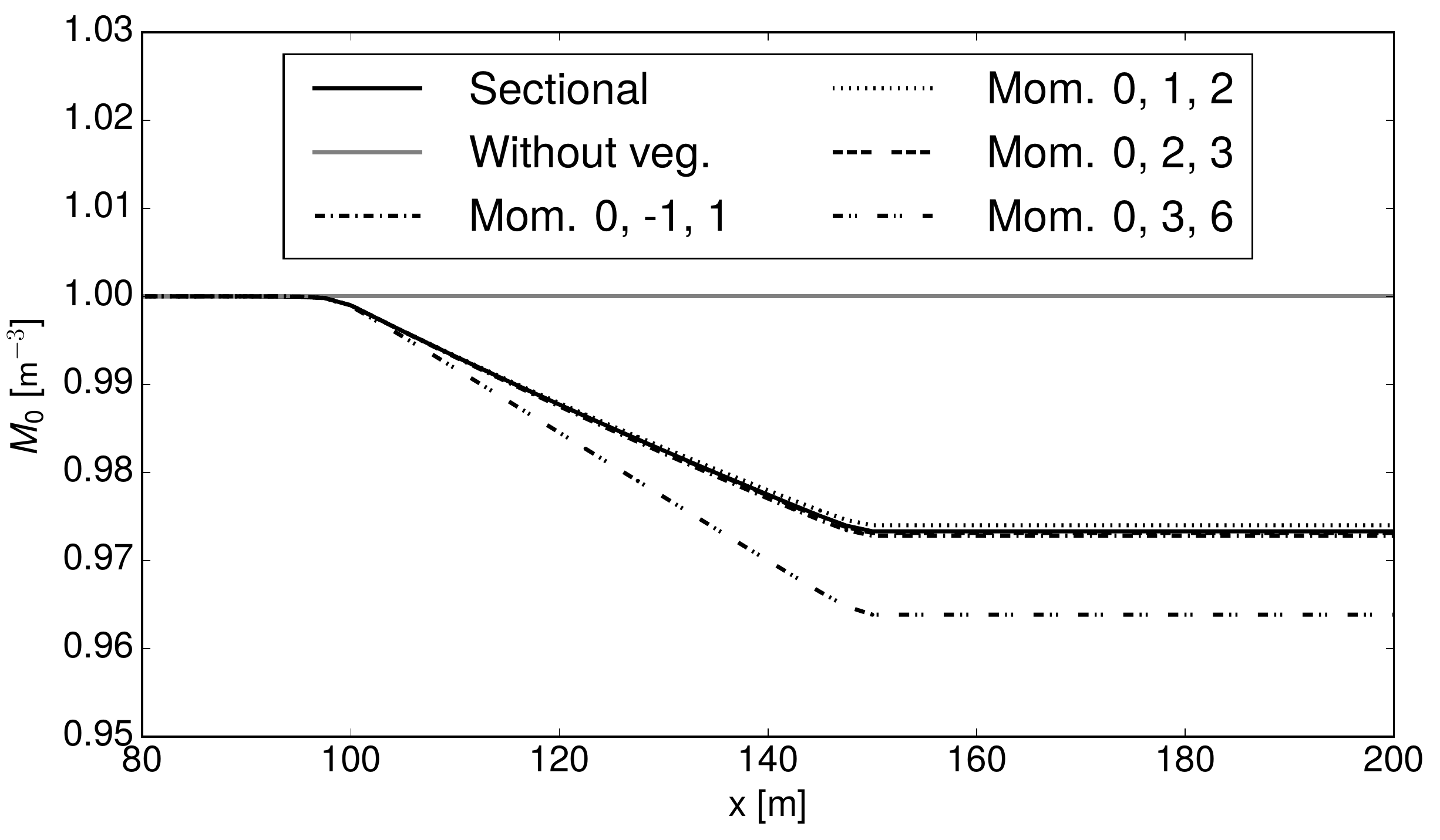}
  \includegraphics[width=0.48\textwidth]{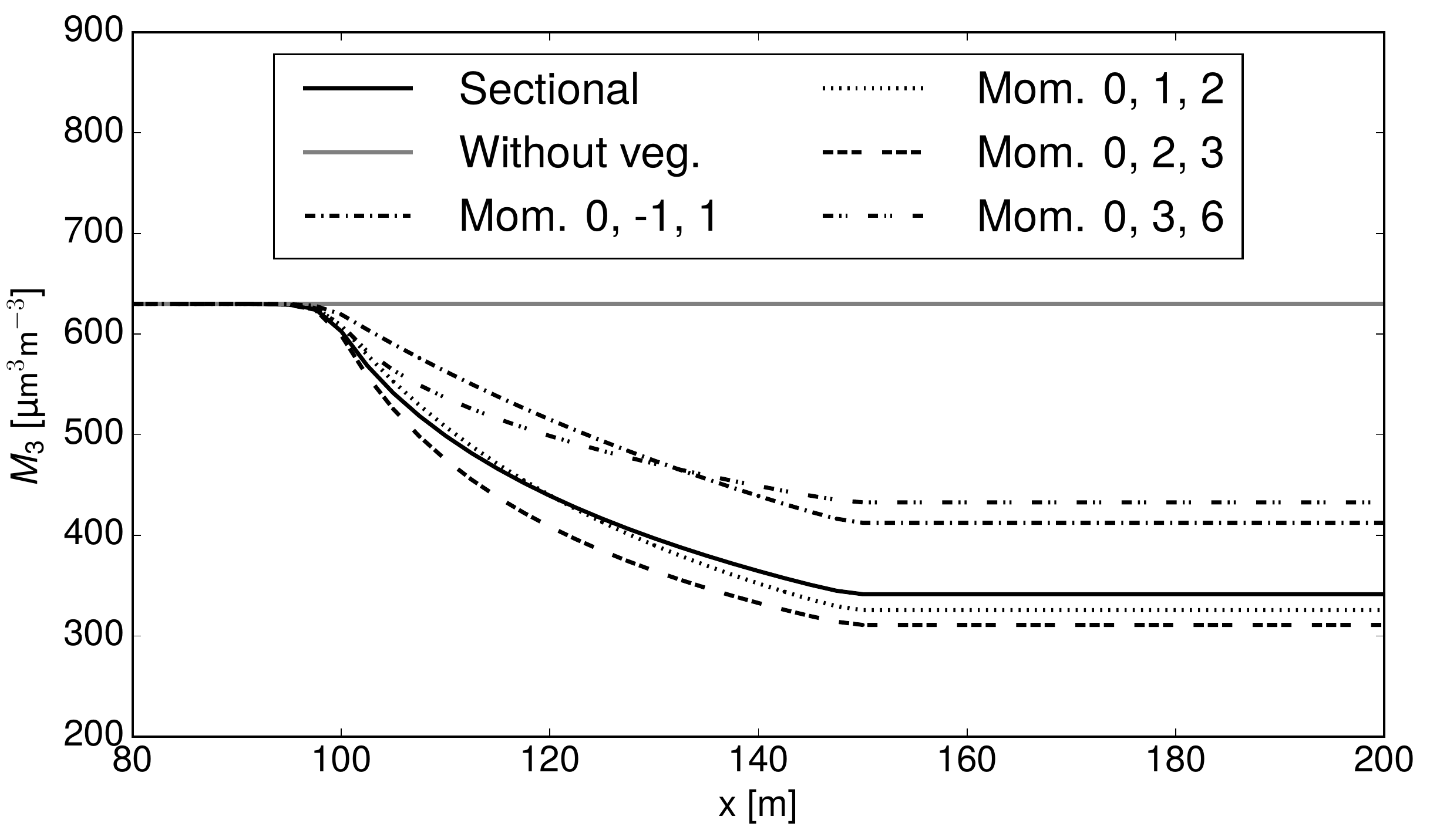}
  \caption{Evolution of the moments along 1-dimensional the tube. (Left) Zeroth moment (Right) Third moment.}
  \label{fig:choice-horiz}
\end{figure}

Number and volume concentration distribution behind the barrier (at 150 m) are shown on Fig. \ref{fig:choice-dist}. As a reference, calculated distributions are complemented by the distribution for a case without the vegetation present. Evolution of the zeroth moment (equal to the number concentration) and the third moment (proportional to the volume concentration) through the vegetation block are shown on Fig. \ref{fig:choice-horiz}.

Effect of the vegetation, while small in number concentration, is significant in volume concentration. Only the variant using the moments of orders 0, -1, and 1 reproduces well the number concentration distribution, but overpredicts the peak of the volume concentration. Variants using the orders 0, 1, 2 and 0, 2, 3 produce result closer to the sectional model in volume concentration, but with larger differences in number concentration. Variant using the orders 0, 3, 6 shows no advantages over the other variants.

Choosing between the orders 0, 1, 2 and 0, 2, 3, we opted for the latter variant, as the third moment is proportionate to the main quantity of interest - volume (and mass) concentration of the pollutant.

\subsection{Numerical implementation}
\label{sec:implementation}

Both the sectional model and the moment model were implemented using the OpenFOAM platform \citep{OpenFoamUserGuide}. Second order upwind scheme was used for convective terms in Equations (\ref{eq:n-eq}) and (\ref{eq:moment-eq}) and second order scheme based on the Gauss theorem was used for the diffusive terms.
Residual levels of $10^{-5}$ were used to test for convergence of the steady state solver.

When using the moment method, we have to solve the discretized Eq. (\ref{eq:moment-eq}) for the three selected moments. These equations are coupled through the gravitational settling term and the deposition term, which depends on the moments of a different order than the one solved by the equation. The coupling is dealt with the following way.
In every iteration, first the parameters of the lognormal distribution $N$, $d_{gn}$ and $\sigma_g$ are calculated using the Equations (\ref{eq:n}-\ref{eq:ln2sigmag}) from the values in the preceeding iteration. Three moment equations are then solved one after another with the coupling terms resulting from the deposition being treated explicitly.

Fully explicit treatment of the gravitational settling term (\ref{eq:mom-grav}) can result in numerical instability, unless low values of the relaxation factors are used. That would however lead to slower convergence, therefore we employed a semi-implicit treatment. Moment $M_{k+2}$ in (\ref{eq:mom-grav}) is rewritten as $M_{k+2} = F_{k,2} M_k$ with
\begin{equation}
F_{k,m} = M_{k+m}/M_k = d_{gn}^m \exp \left( \frac{m (m + 2k)}{2} \ln^2 \sigma_g \right)
\end{equation}
and the term $F_{k,2}$ is then treated explicitly and $M_k$ implicitly.

Relaxation factors 0.95 were used both for the sectional equations and for the moment equations. For the first five iterations of the moment method the relaxation factors for the moment equations were however set to lower value 0.8, as the computations proved to be less stable at the beginning.

Calculation of the distribution parameters $d_{gn}$ and $\ln^2 \sigma_g$ via Eq. (\ref{eq:dgn}) and (\ref{eq:ln2sigmag}) includes the division of the moments, potentially very small far away from the source of pollutant. To avoid this problem, small background concentration in the whole domain is set as an initial condition and used as a boundary condition where zero would be used otherwise.

\section{Applications}

Here we describe two example problems of microscale flows through and around the vegetation and assess the applicability of the developed moment method to the simulation of pollutant dispersion. Two vegetation elements that could be encountered in the urban settings are investigated in this test: small patch of full grown trees and a dense hedgerow.

The flow field in both cases was precomputed by an in-house finite volume CFD solver. The solver is based on the Navier-Stokes equations in the Boussinesq approximation and utilizes $k-\epsilon$ turbulence model.
Inlet profiles of velocity and the turbulence quantities, as well as the wall functions, are prescribed by the analytical expressions given by \citet{RichardsHoxey93}.
Vegetation model for the momentum and $k-\epsilon$ equations described by \citet{Katul04} is employed.
For further details we refer to \citep{SipBenes15b}, where the solver is described in more detail.
Turbulent Schmidt number was set to $Sc_T = 0.7$ in both cases, based on the analysis by \citet{TominagaStathopoulos07}.

In both cases presented below, we simulated the dispersion of a coarse mode particles from a point or a line source.
The coarse mode is chosen as the mode that contains, together with the accumulation mode, majority of the volume of the particles in the urban environment \citep{SeinfeldPandis06}, but is affected more strongly by the dry deposition than the accumulation mode.
The number distribution at the source is assumed to be lognormal with the parameters $d_{gn} = 0.86 \si{\micro \meter}$ and $\sigma_g = 2.21$, typical for the urban environment \citep{Hinds99}.

Evaluation of the developed moment method was based on the comparison with the results obtained by the sectional model.
In the sectional model, Eq. (\ref{eq:n-eq}) is solved for 41 particle sizes distributed uniformly between \SI{0.01}{\um} and \SI{100}{\um}.
The interval is chosen so that the behaviour of the number distribution as well as the volume distribution can be captured by the sectional model.

\subsection{Tree patch in 2D}
\label{sec:veg2d}

First case investigates the filtering properties of a small patch of full grown conifer trees. A simplified 2D model is constructed as follows. The 30 meters wide and 15 meters high tree patch is represented as a horizontally homogeneous vegetation block. Pollutant source is placed 15 meters upstream from the vegetation, 5 meters above the ground.
\begin{figure}[h]
  \centering
  \raisebox{0.1\height}{\includegraphics[width=0.5\textwidth]{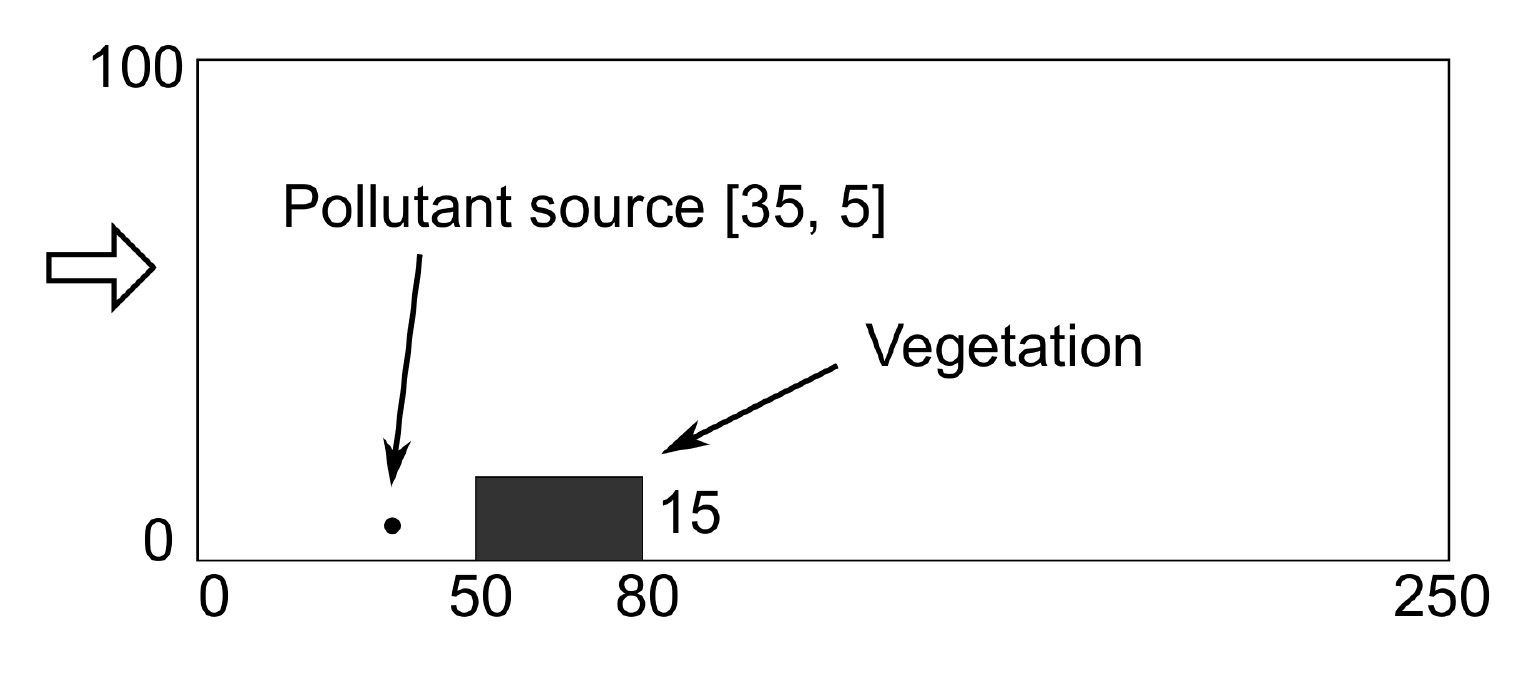}}
  \includegraphics[width=0.3\textwidth]{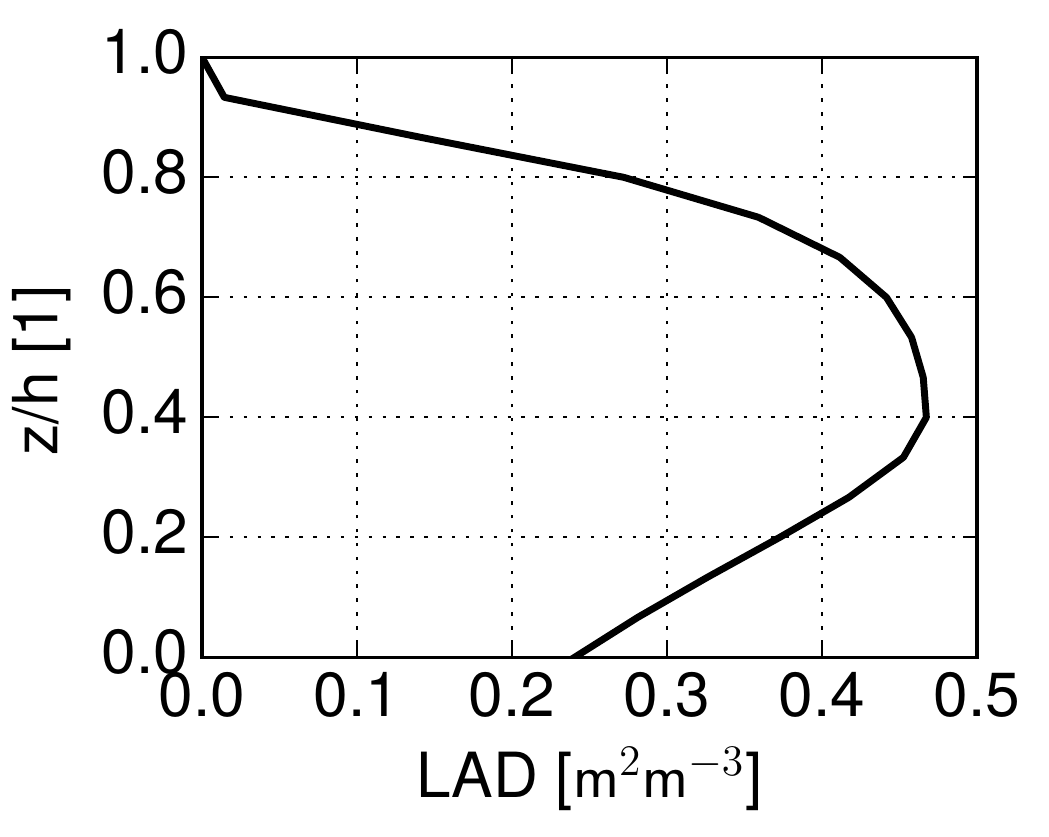}
  \caption{(Left) Sketch of the domain. All dimensions in meters. Sketch is not to scale. (Right) LAD profile of the vegetation.}
  \label{fig:veg2d-domain}
\end{figure}
LAD profile of the vegetation is prescribed by a formula given by \citet{LalicMihailovic04},
\begin{align}
  \mathrm{LAD}(z)& = L_m \left(\frac{h - z_m}{h - z} \right)^n \exp\left(n\left(1 - \frac{h - z_m}{h - z}\right)\right), \\
  n   & =
  \left\{
    \begin{array}{ll}
      6   & \mbox{if } 0 \leq z < z_m, \\
      0.5 & \mbox{if } z_m \leq < z \leq h,
    \end{array}
  \right. \nonumber
\end{align}
where $h = 15$ m is the height of the trees, $L_m$ is the maximum LAD, chosen so that leaf area index, ${\mathrm{LAI} = \int_0^h \mathrm{LAD}(z) \diff z}$, is equal to 5, and $z_m = 0.4 h$ is the corresponding height of maximal LAD.
The sketch of the domain and the LAD profile of the vegetation is shown on Fig. \ref{fig:veg2d-domain}.
Trees are modelled as generic conifers with $d_e = \SI{2}{\mm}$. The drag coefficient is chosen as $C_d = 0.3$ \citep{Katul04}.

Intensity of the point source is set to a normalized value \SI{1}{\per\s} in terms of number of particles. Since all terms in Eq. (\ref{eq:n-eq}) and Eq. (\ref{eq:moment-eq}) are linear with respect to the number concentration, results can be simply scaled to other value of the source intensity if needed.

Inlet wind profile is set as logarithmic with $u_{\mathrm{ref}} = \SI{10}{\m\per\s}$ at height 20~m and $z_0 = 0.1$~m.
For the number concentration in the sectional model and for all moments in the moment method the Neumann boundary conditions are used on the ground, at the top and at the outlet.
No resuspension of the particles is allowed, i.e. any particle that falls on the ground stays on the ground indefinitely.
Small value of the concentration and of the moments calculated from the lognormal distribution with the parameters $N = 10^{-6} \si{\m^{-3}}, d_{gn} = \SI{0.86}{\um}, \sigma_g = 2.21$ is prescribed at the inlet.

Domain is discretized using a cartesian grid with 220 cells in horizontal direction and 100 cells in vertical direction, graded so that the grid is finer near the ground and around the tree patch. The near ground cells are 0.25~m high, and the vegetation block itself consists of 42 x 40 cells.

\begin{figure}[h]
  \centering
  \includegraphics[width=0.7\textwidth]{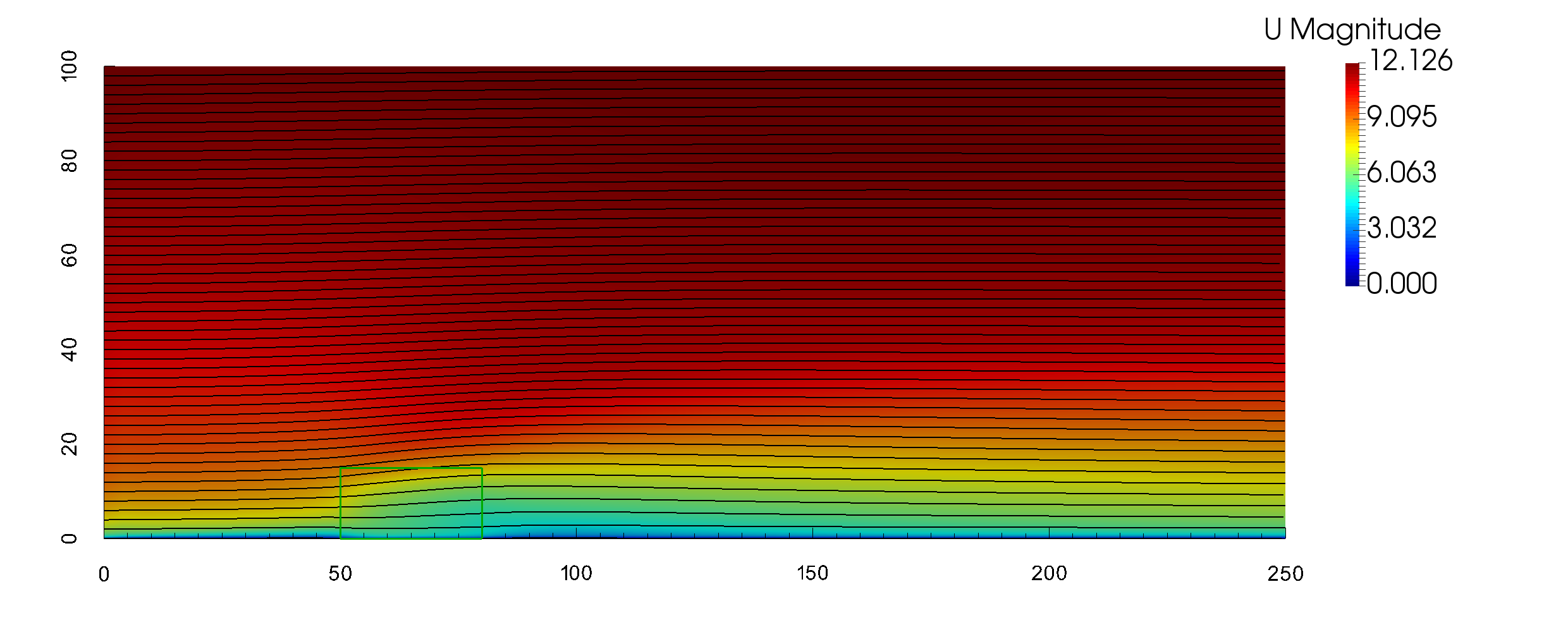}
  \caption{Flow field for the 2D tree patch case. Shown are the streamlines, background is coloured by velocity magnitude (in \si{\m\per\s}). Position of the tree patch is marked by a green rectangle.}
  \label{fig:veg2d-streamlines}
\end{figure}

Flow field obtained by the CFD solver is shown on Fig. \ref{fig:veg2d-streamlines}.
As visible, the vegetation block slows the wind down, but allows the air to pass through.

Results from the sectional and the moment model are compared in terms of the third moment of the particle size distribution, proportionate to the volume concentration of the particles. As we assume that the density is the same for particles of all sizes, third moment is also proportionate to the mass concentration of the particles.

\begin{figure}[h]
  \centering
  \includegraphics[width=0.7\textwidth]{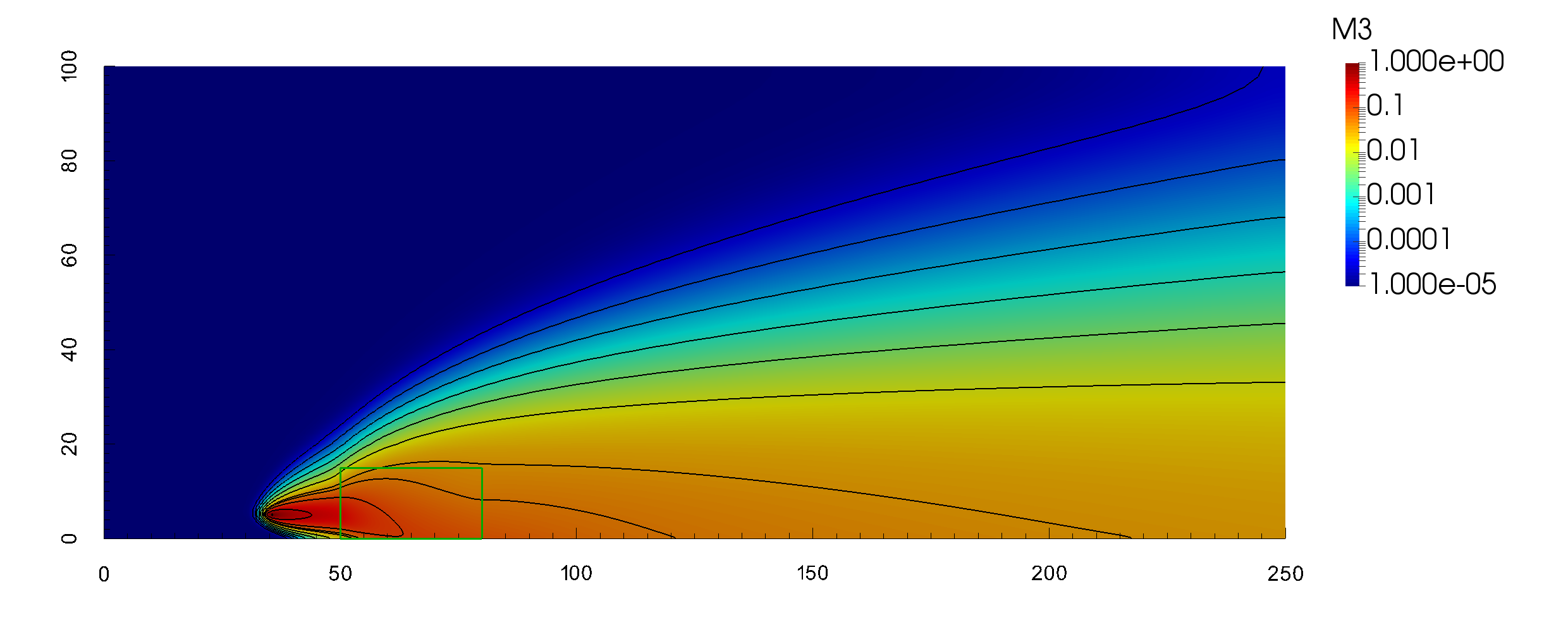}
  \includegraphics[width=0.7\textwidth]{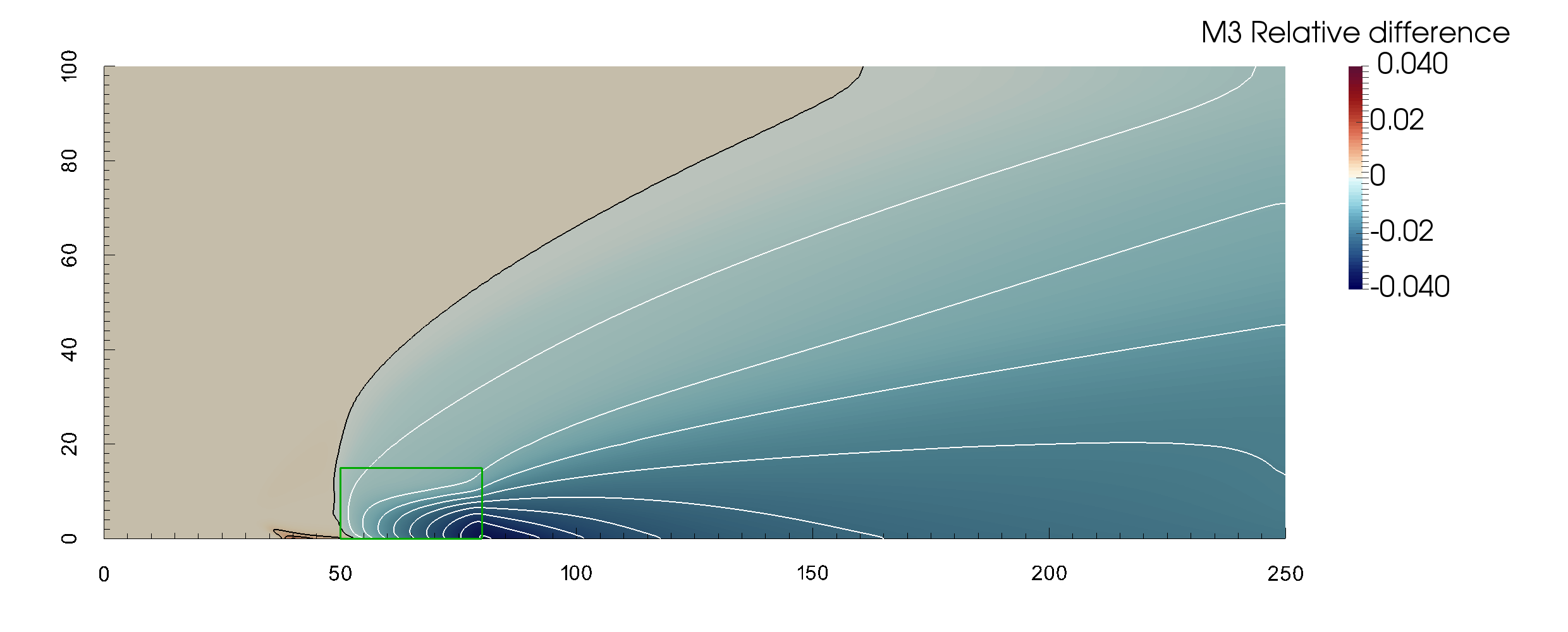}
  \caption{Results for the 2D tree patch case. (Top) Third moment of the size distribution calculated by the moment method (in \si{\um^3 \m^{-3}}). (Bottom) Relative difference $(M_3^{mm} - M_3^{sec})/M_3^{mm}$ of the third moment calculated by the moment method and the sectional approach.}
  \label{fig:veg2d-M3}
\end{figure}

Calculated field of the third moment by the moment method is shown on Fig. \ref{fig:veg2d-M3} (left). The relative difference $(M_3^{mm} - M_3^{sec})/M_3^{mm}$ of the results obtained by the moment method, $M_3^{mm}$, and by the sectional model, $M_3^{sec}$, is shown on the right panel of Fig. \ref{fig:veg2d-M3}.

The source of the largest discrepancies between the two methods is the vegetation block. The relative difference raises up to 4\% at the downstream edge of the vegetation block, and then decreases with the increasing distance from the vegetation.
The moment method overpredicts the deposition inside the vegetation due to the inexact approximation of the impaction efficiency described in section \ref{sec:vd-IM}.

\begin{figure}[h]
  \centering
  \includegraphics[width=0.4\textwidth]{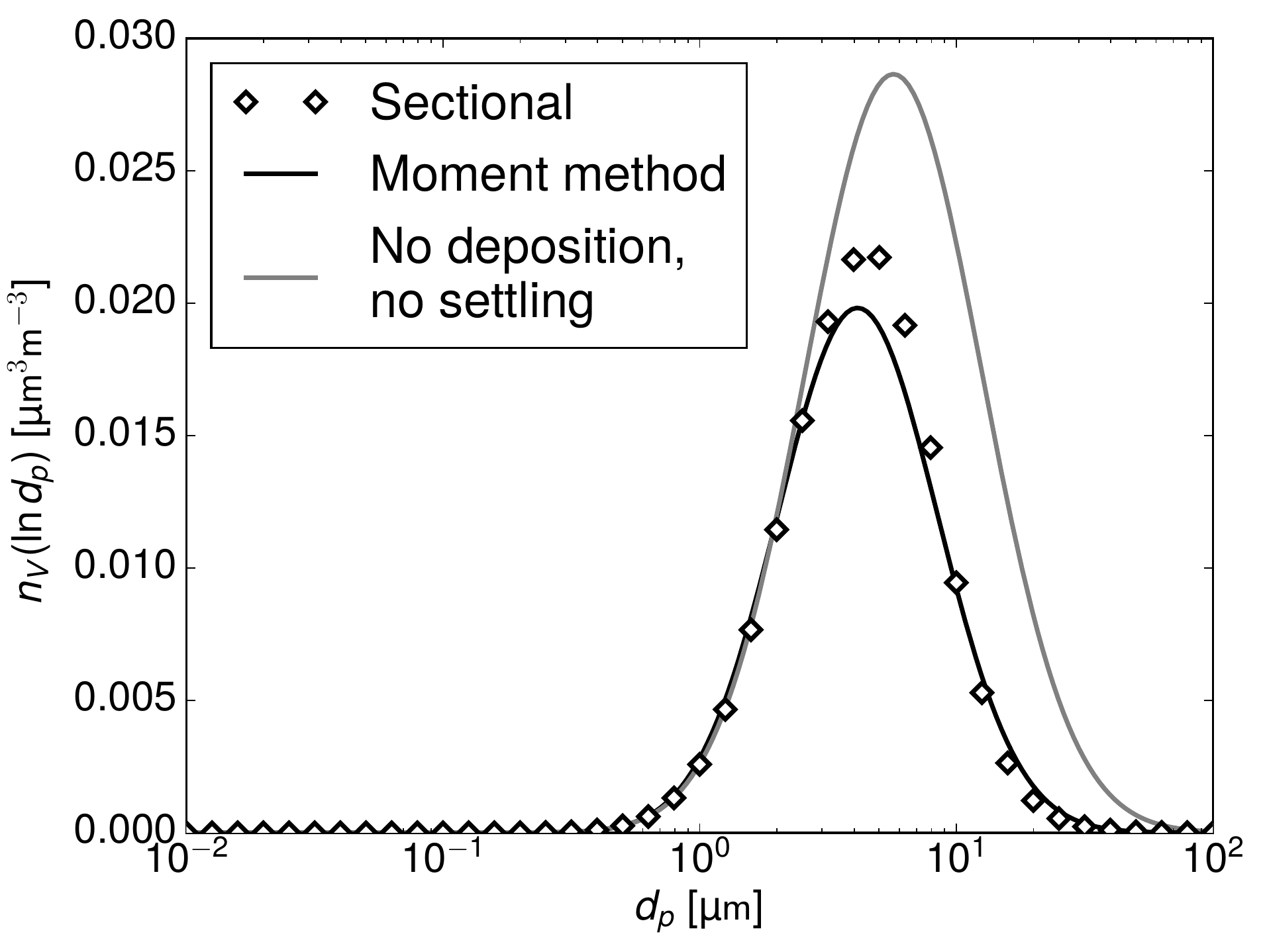}
  \includegraphics[width=0.4\textwidth]{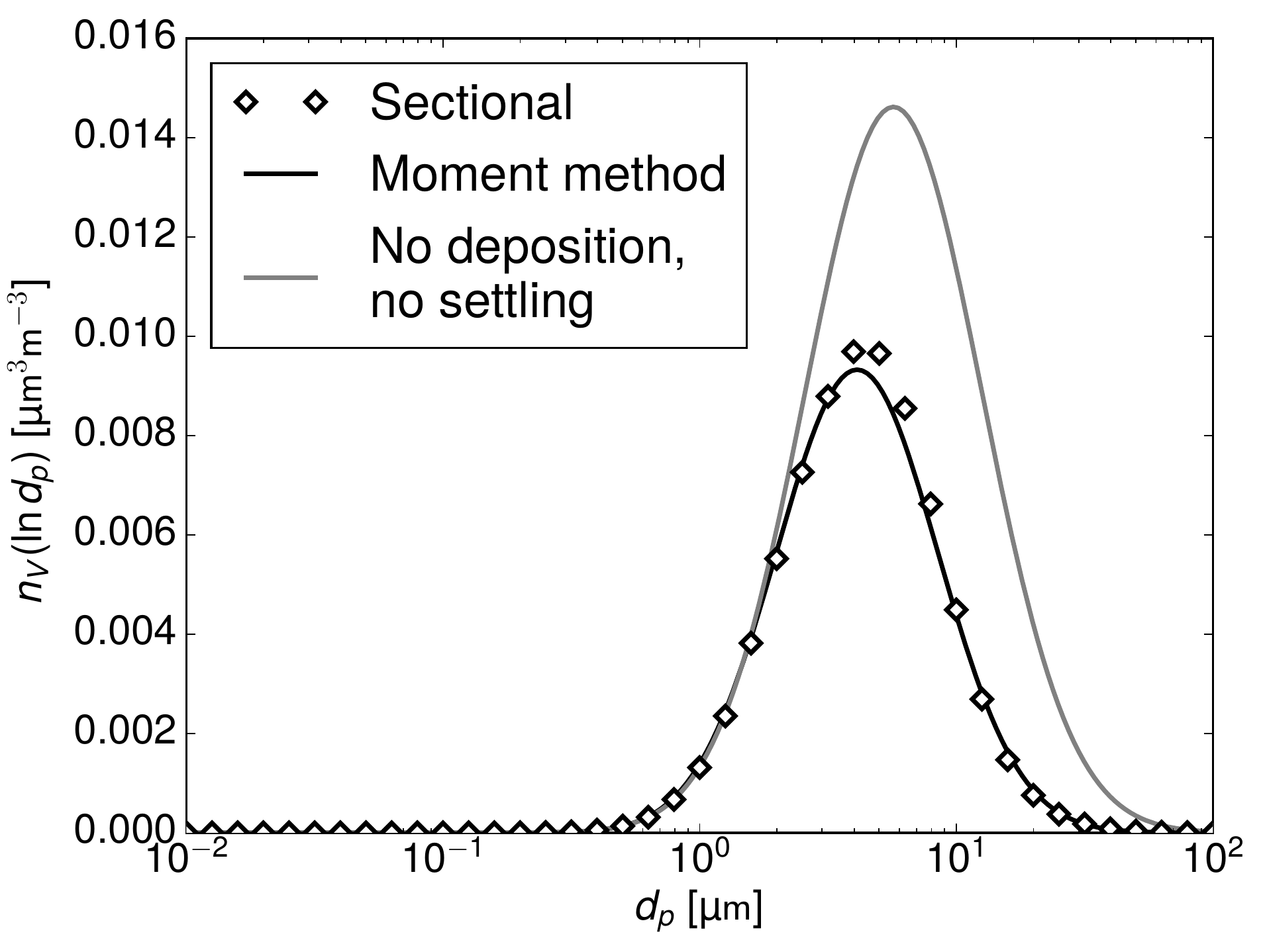}
  \caption{Results for the 2D tree patch case. (Left) Volume concentration at [80; 2] (Right) Volume concentration at [200; 2]. Discrete points calculated by the sectional method and the distribution calculated by the moment method are shown. For reference, the distribution calculated without the size dependent deposition and gravitational settling terms is shown as well.}
  \label{fig:veg2d-dist}
\end{figure}

Further insights can be obtained from Fig. \ref{fig:veg2d-dist}. It shows the volume concentration distribution at the downstream edge of the tree patch, and at 120~m downstream from the tree patch, both at height 2 m above ground.
The vegetation has negligible effect on the particles smaller than \SI{2}{\micro\meter}, but significantly reduces the mass of the particles above \SI{10}{\micro\meter}. This is captured well both by the sectional and the moment method.
At the downstream edge of the tree patch the moment method predicts lower peak of the volume concentration distribution than the sectional approach. The difference is reduced by the mixing of the filtered air with the unfiltered air flowing above the vegetation further away from the tree patch.

\subsection{Hedgerow in 3D}
\label{sec:case3d}

Next we tested the method on a 3D model of a dense hedgerow placed near a line source of the pollutant. This case is a three dimensional extension of the 2D situation investigated in \citep{TiwaryEtAl05}.
The yew hedge is 10~m wide, 3.2~m deep and 2.4~m high. It is placed in the 40~m wide, 40~m long, and 20~m high computational domain.
Two meters upstream from the hedge is a line source at height 0.5~m above ground.
Intensity of the line source is set to a value \SI{1}{\per\s\per\m} in terms of number of particles, noting as in section \ref{sec:veg2d} that the results can be scaled if other value is desired.

\begin{figure}[h]
  \centering
  \includegraphics[width=0.50\textwidth]{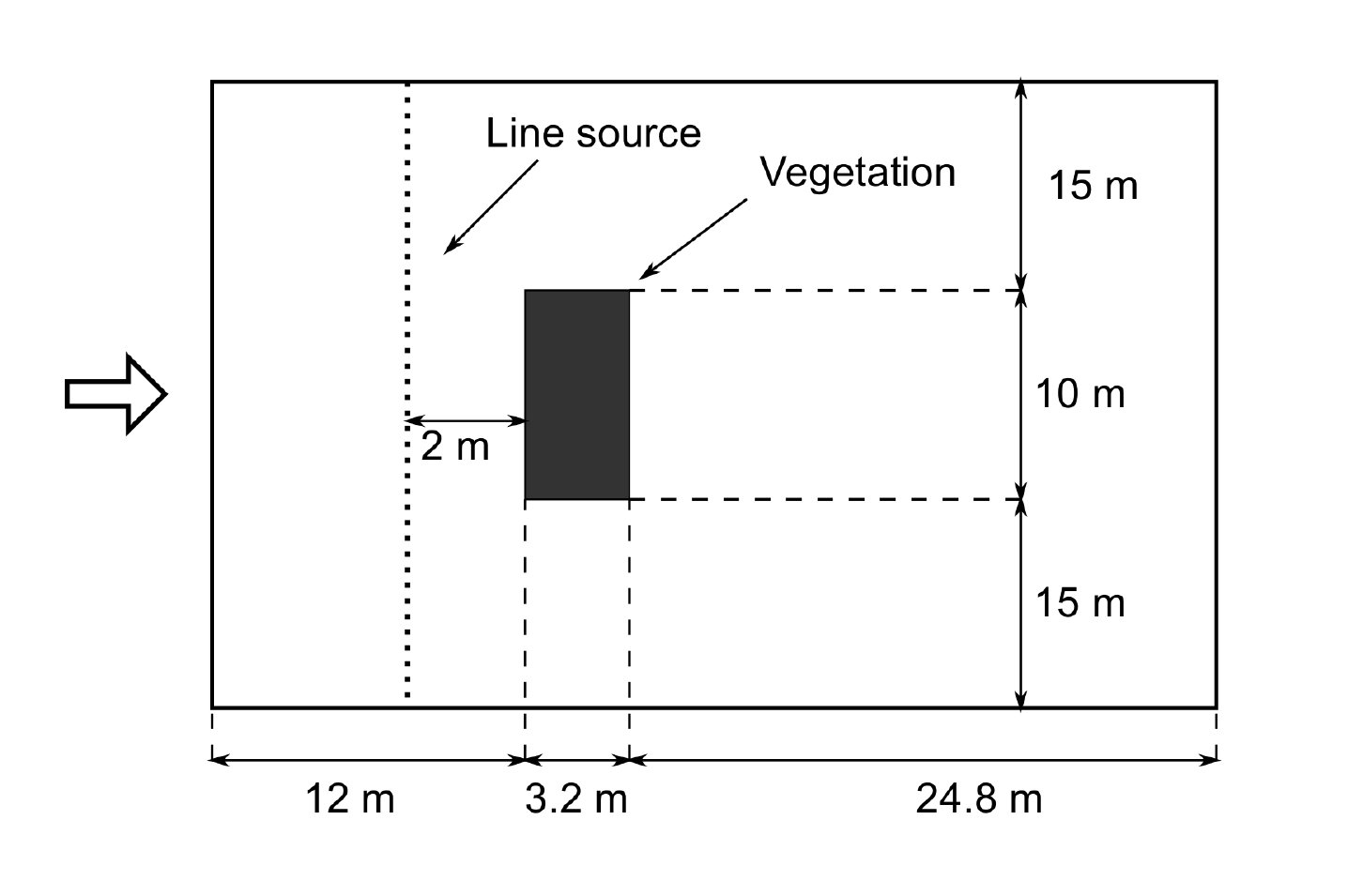}
  \raisebox{0.2\height}{\includegraphics[width=0.3\textwidth]{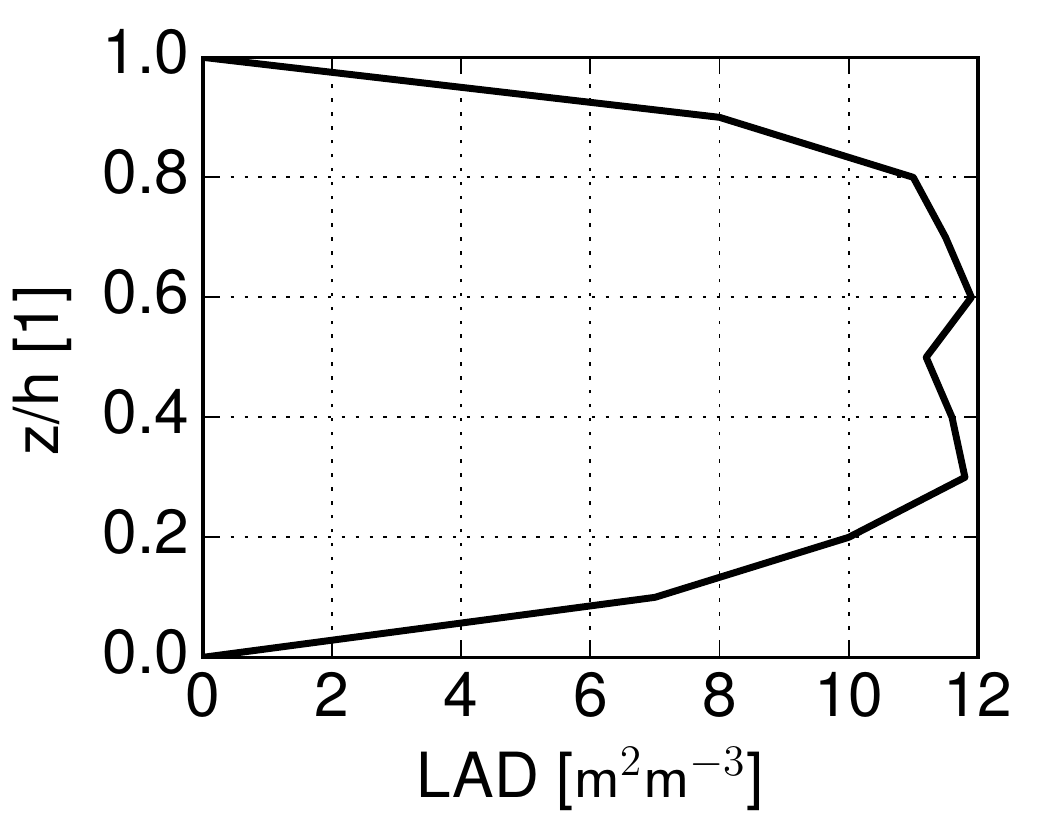}}
  \caption{(Left) Overhead view of the domain (not to scale) (Right) LAD profile of the vegetation.}
  \label{fig:veg3d-domain}
\end{figure}

Sketch of the domain is shown on the left panel of Fig. \ref{fig:veg3d-domain}. Right panel shows the LAD profile of the hedge, taken from the original article.
Vegetation is further described by the needle diameter is, $d_e = \SI{3}{\mm}$, and the vegetation drag coefficient which is set to $C_d = 0.5$ as in \citep{TiwaryEtAl05}.

The computational mesh was created using the OpenFOAM \textit{snappyHexMesh} generator. The domain consist of 376 000 cells, refined near the ground and around the hedge. The near-ground cells are 0.07~m high and the hedge itself is discretized using 54 x 20 x 22 cells.

Wind profile at the inlet is set as logarithmic with $u_{\mathrm{ref}} = \SI{2.5}{\m\per\s}$ at height 2.4~m and $z_0 = 0.1$~m.
Boundary conditions for the sectional solver and moment method solvers are set similarly as in section \ref{sec:veg2d}: Neumann boundary conditions are used at the ground, top, sides, and at the outlet. No resuspension of the particles fallen to the ground is allowed.
Again, small amount of the particles given by the lognormal distribution with the parameters $N = 10^{-6}, d_{gn} = \SI{0.86}{\um}, \sigma_g = 2.21$ is prescribed at the inlet.

\begin{figure}[h]
  \centering
  \includegraphics[width=1.0\textwidth]{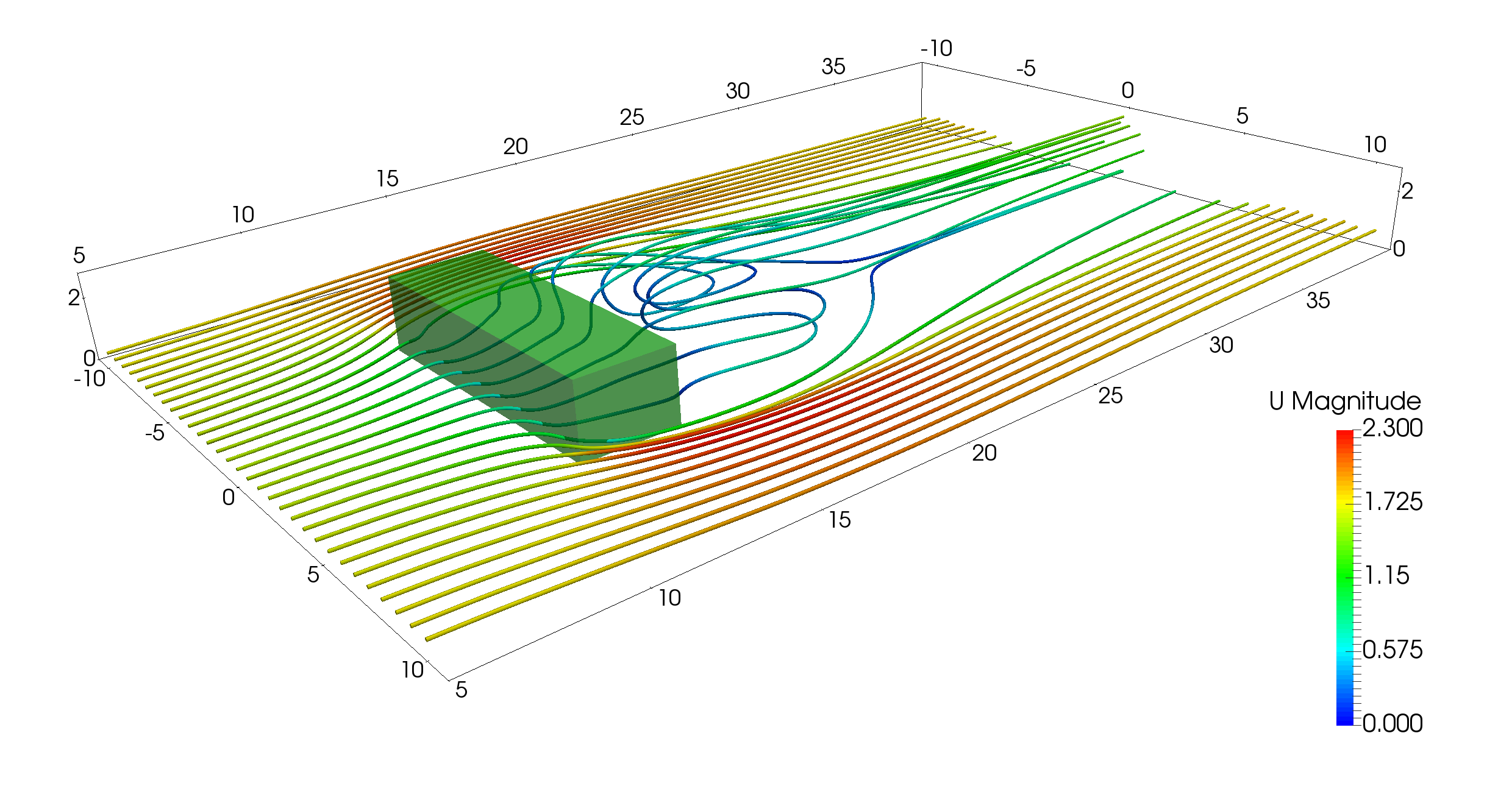}
  \caption{Streamlines of the flow around the hedgerow. Streamlines are released at height 0.5~m and are coloured by the velocity magnitude.}
  \label{fig:veg3d-streamlines}
\end{figure}

Streamlines of the flow field calculated by the separate CFD solver are shown on Fig. \ref{fig:veg3d-streamlines}. As in the 2D simulation in \citep{TiwaryEtAl05}, recirculation zone is developed behind the dense hedge. Unlike the 2D case, here we can observe part the of the flow to be deflected to the sides.

\begin{figure}[h]
  \centering
  \includegraphics[width=0.7\textwidth]{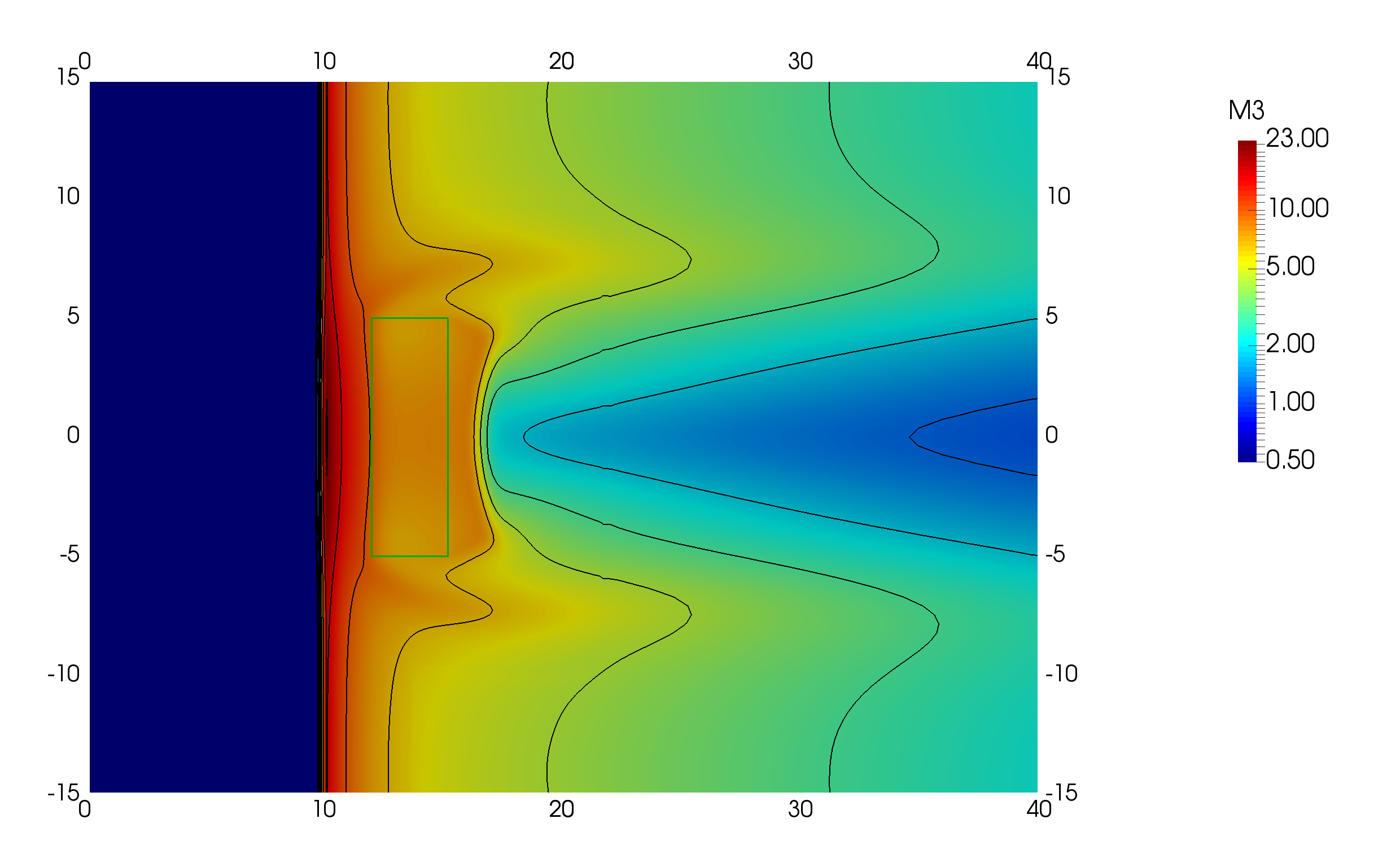}
  \includegraphics[width=0.7\textwidth]{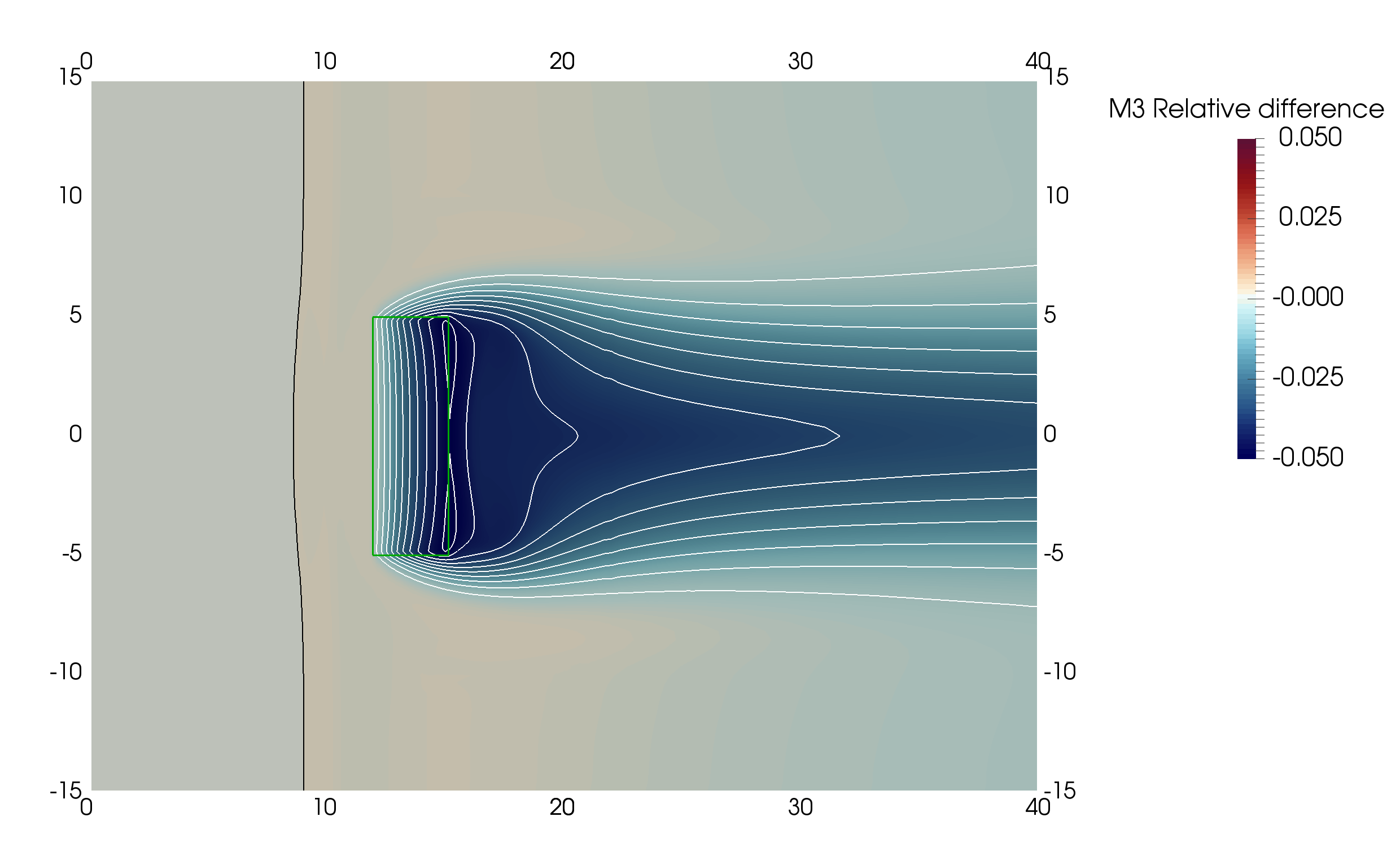}
  \caption{Results for the 3D hedgerow case. Horizontal cut at height $z = 0.5$~m. Quantities shown are as on Fig.~\ref{fig:veg2d-M3}.}
  \label{fig:veg3d-M3-horiz}
\end{figure}

\begin{figure}[h]
  \centering
  \includegraphics[width=0.7\textwidth]{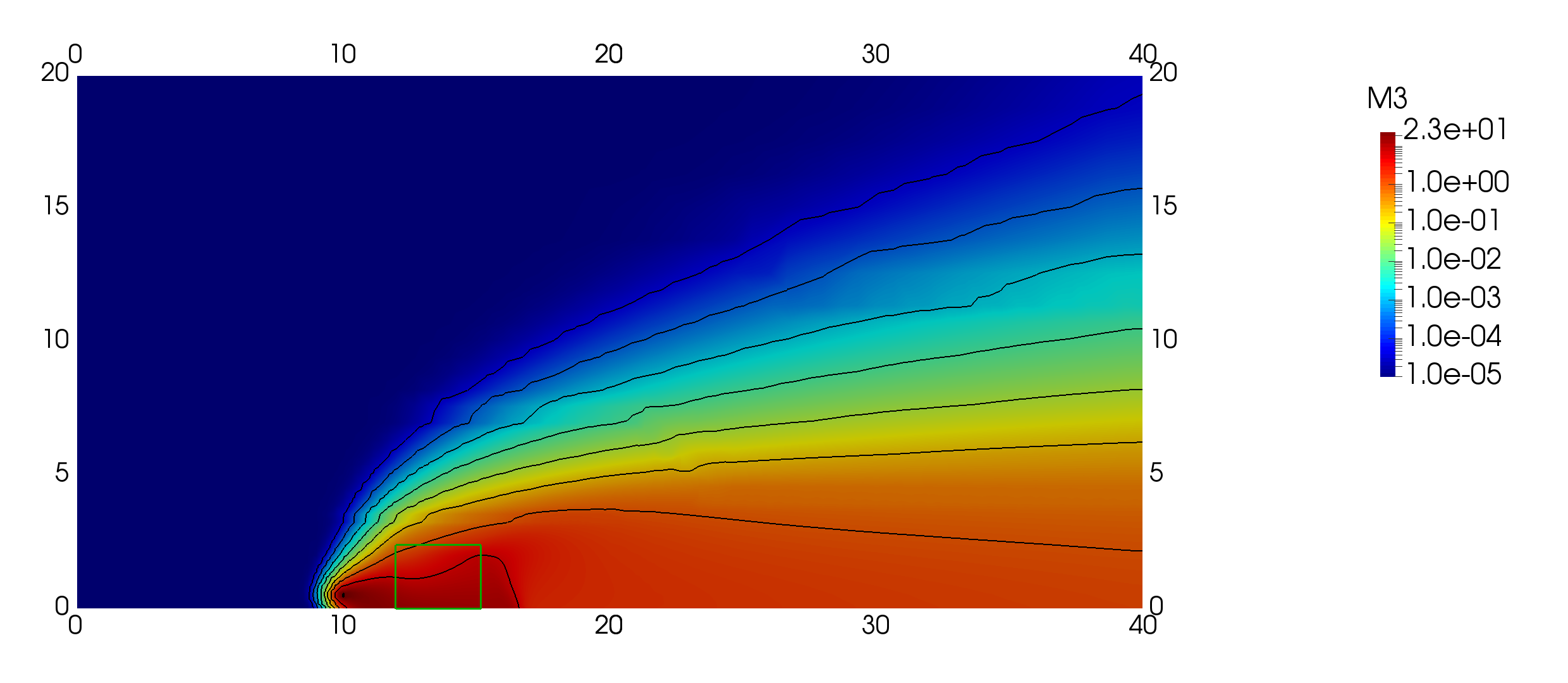}
  \includegraphics[width=0.7\textwidth]{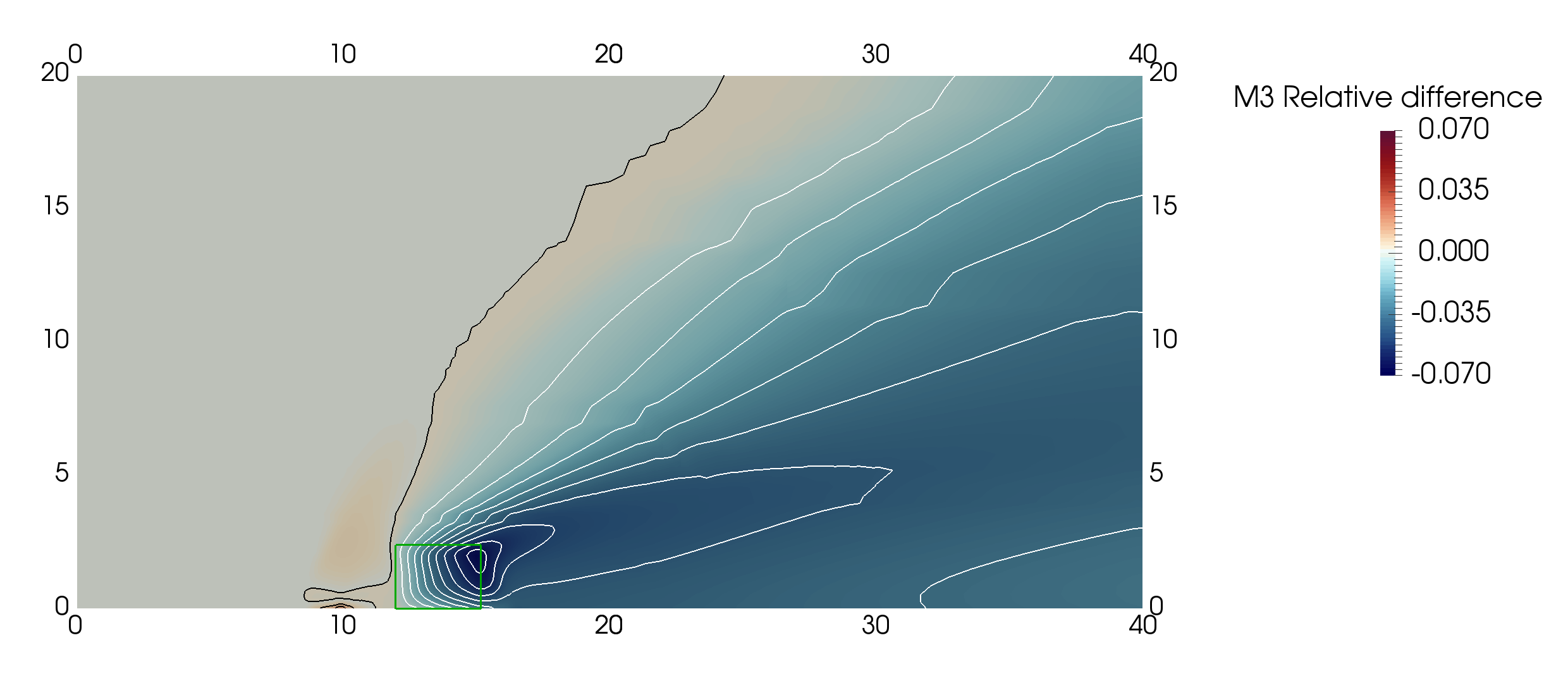}
  \caption{Results for the 3D hedgerow case. Vertical cut at $y = 0$~m. Quantities shown are as on Fig.~\ref{fig:veg2d-M3}.}
  \label{fig:veg3d-M3-vert}
\end{figure}

Third moment of the particle size distribution obtained by the moment method is shown on the left panels of Fig. \ref{fig:veg3d-M3-horiz} and Fig. \ref{fig:veg3d-M3-vert}. While a portion of the pollutant penetrates the barrier, part is deflected to the sides of the hedgerow, creating a zone with a reduced pollutant concentration behind it.

Relative difference between the solution obtained by the moment method and sectional approach is shown on the right panels of  Fig. \ref{fig:veg3d-M3-horiz} and Fig. \ref{fig:veg3d-M3-vert}. As in the tree patch case, the moment method overpredicts the deposition and consequently underestimates the volume concentration behind the barrier. The difference is below 10\%, and decreases away from the barrier.

Effects of the coarser mesh in the upper part of the computational domain are visible on Fig. \ref{fig:veg3d-M3-vert}. However, it does not negatively affect the difference between the two methods.

\begin{figure}[h]
  \centering
  \includegraphics[width=0.4\textwidth]{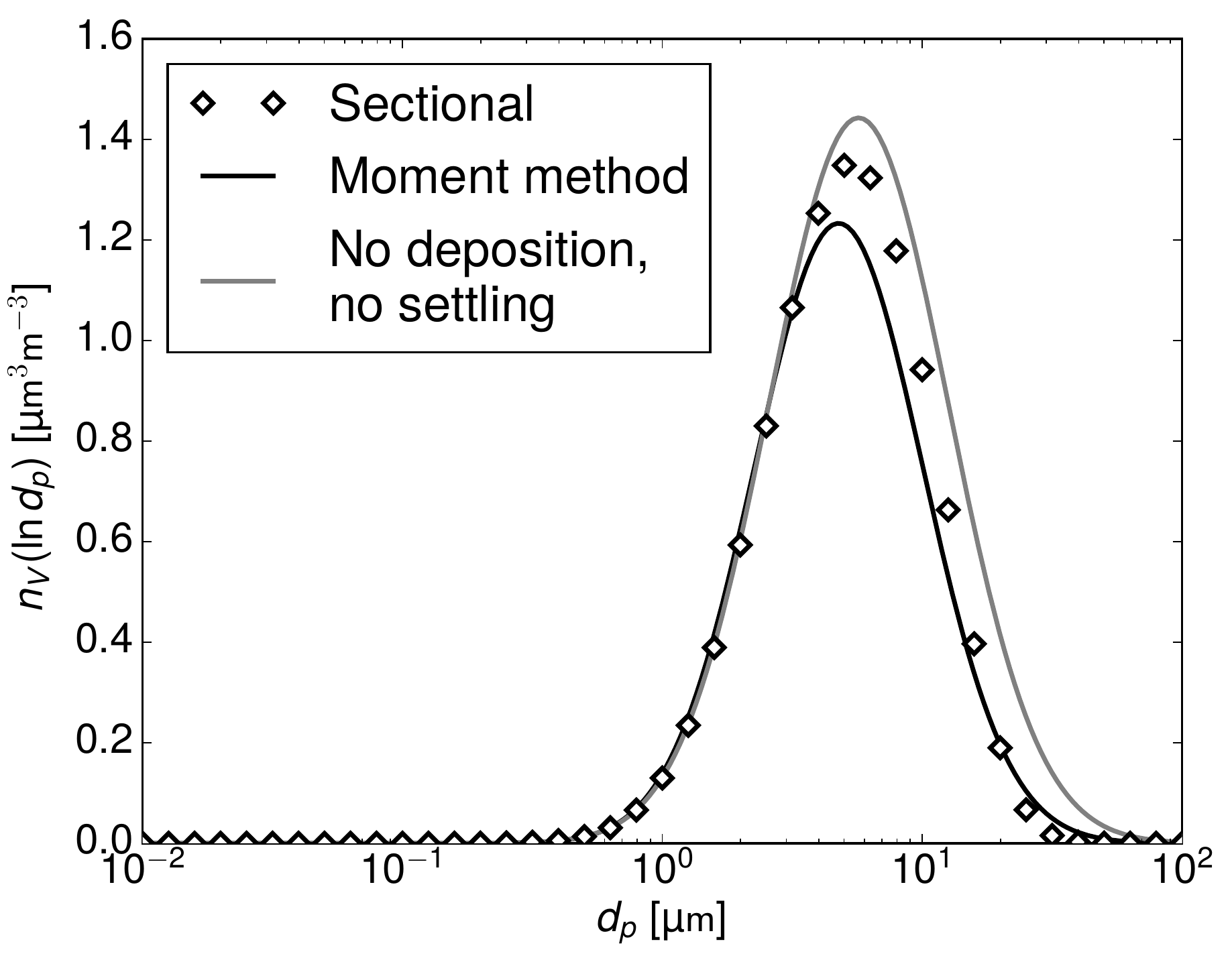}
  \includegraphics[width=0.4\textwidth]{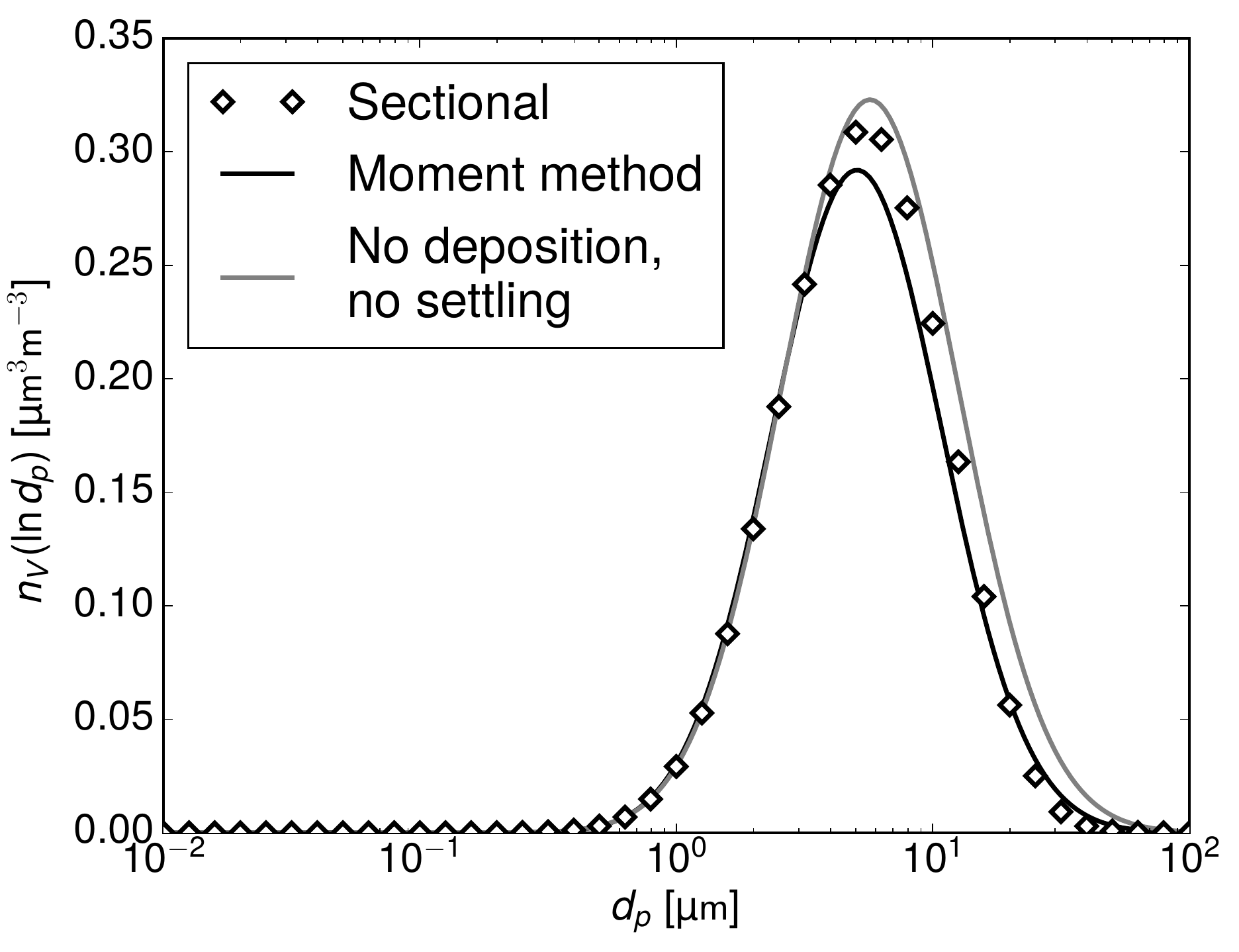}
  \caption{Results for the 3D hedgerow case. (Left) Volume concentration at [15; 0; 2]. (Right) Volume concentration at [30; 0; 2].
    For reference, the distribution calculated without the size dependent deposition and gravitational settling terms is shown.}
  \label{fig:veg3d-dist}
\end{figure}

Volume concentration distribution at two points - inside the vegetation and downstream from the vegetation - is shown on Fig. \ref{fig:veg3d-dist}.
Due to the smaller size of the vegetation than in the 2D tree patch case, the effect of the vegetation is less pronounced.
The moment method is able to reproduce the shape of the distribution well, but again produces a lower peak than the sectional method. Similarly as before, better fit can be observed further from the barrier due to the mixing with unfiltered air.

\subsection{Computational performance}

To compare the computational performance of the developed model, we measured the runtime of the sectional approach and the moment method approach for the 3D case described in section \ref{sec:case3d}. Both solvers were run on a single core of an Intel Xeon X5365 processor.

The sectional model, comprised of 41 scalar PDEs, finished in 9120 seconds. The average runtime per each equation was thus 222 seconds.
Moment model, comprised of 3 coupled PDEs, finished in 1128 seconds. That gives us eightfold acceleration compared to the sectional model.
Even though the high number of bins used in this study might not be necessary to obtain sufficiently accurate results, to get an equivalent workload as the moment method in this case, only 5 bins could be used in the sectional model. Such number is insufficient to model the behaviour of the number distribution as well as the volume distribution well.

Two points can be made in favor of the sectional method though.
First, the solution process of every equation is independent on the other equations, therefore the approach offers effortless parallelization for the number of cores up to the numbers of bins used. This is not especially advantageous in our implementation, as the OpenFOAM solvers are already parallelizable, but it could be an important factor for other implementations.
Secondly, the relaxation factor 0.95 used for all simulations in the sectional approach was needed only for the bins representing the larger particles. Using different values of this parameter for different bins can provide some reduction of the runtime.

\section{Conclusions}

In this study, we introduced a formulation of a dry deposition model suitable for implementation in a moment method. As the original model by \citet{PetroffEtAl08b}, our approximation includes five main processes of the dry deposition: Brownian diffusion, interception, impaction, turbulent impaction, and sedimentation.

The developed deposition velocity model was implemented in a microscale finite volume solver based on the OpenFOAM platform. The solver employs the moment method to calculate the particle size distribution in the domain.
The deposition model was tested on two example problems of microscale pollutant dispersion. Comparison with the sectional method using the original dry deposition model revealed that the moment method is able to reproduce the shape of the particle size distribution well. The relative differences between the sectional and the moment method in terms of the third moment of the distribution were below 10 \%.
This difference was caused by the higher deposition velocity in our model, which in turn was caused by the inexact representation of the impaction process.
The impaction term could be represented better by employing numerical integration, that would however result in a higher computational costs, and reduce the advantages of using the moment method.

The moment method, described by three coupled PDEs, proved to be more computationally efficient than the sectional model using 41 bins. The speedup was approximately eightfold, and a workload equivalent to the moment method would be achieved by running a sectional model that uses only 5 bins.

This performance improvement together with the reliable results shows that the moment methods, often used in large scale atmospheric models, can be useful also for the microscale problems of pollutant dispersion in the urban environment.

The developed method as formulated here is applicable only when the particle size distribution can be approximated as a lognormal distribution. Here we used only unimodal distribution, but the usage of multimodal distribution would be also possible by superposition of several unimodal ones.
Furthermore, the method could be reformulated for other distributions, provided that algebraic relations between the moments and distribution parameters are known.

\section*{Acknowledgements}

This work was supported by the grant SGS16/206/OHK2/3T/12 of the Czech Technical University in~Prague.
Authors would like to thank to Jan Halama and Jiří Fürst for helpful discussions.

\nomenclature{$\bs{u}$}{Wind velocity [\si{\m\per\s}]}
\nomenclature{$U$}{Wind velocity magnitude [\si{\m\per\s}]}
\nomenclature{$u_f$}{Local friction velocity [\si{\m\per\s}]}

\nomenclature{$\bs{u}_s (d_p)$}{Particle settling velocity [\si{\meter\ \second^{-1}}]}
\nomenclature{$u_d (d_p)$}{Deposition velocity [\si{\meter \second^{-1}}]}

\nomenclature{LAD}{Leaf Area Density [\si{\meter^2 \meter^{-3}}]}
\nomenclature{$C_C$}{Cunningham correction factor [1]}
\nomenclature{$\bs{g}$}{Gravity acceleration vector [\si{\meter \second^{-2}}]}

\nomenclature{$D$}{Diffusion coefficient, $D = \nu_T/\mathrm{Sc}$ [\si{\m\cubed\per\second}]}
\nomenclature{$\nu_T$}{Turbulent kinematic viscosity [\si{\m\cubed\per\second}]}
\nomenclature{$Sc_T$}{Turbulent Schmidt number [1]}

\nomenclature{$T_a$}{Temperature of the air [K]}
\nomenclature{$\mu_a$}{Dynamic viscosity of the air [\si{\kilogram\ \meter^{-1} \second^{-1}}]}
\nomenclature{$\nu_a$}{Kinematic viscosity of the air [\si{\meter^{2} \second^{-1}}]}
\nomenclature{$\rho_a$}{Air density [\si{\kilogram \ \meter^{-3}}]}

\nomenclature{$n (d_p)$}{Number concentration of particles [\si{\meter^{-3}}]}
\nomenclature{$n_V (d_p)$}{Volume concentration of particles [\si{\um^{3} \meter^{-3}}]}
\nomenclature{$M_k$}{$k$-th moment of the size distribution [\si{\meter^k \meter^{-3}}]}

\nomenclature{$N$}{Total number concentration [\si{\meter^{-3}}]}
\nomenclature{$S$}{Total surface concentration [\si{\um^2 \meter^{-3}}]}
\nomenclature{$V$}{Total volume concentration [\si{\um^3 \meter^{-3}}]}

\nomenclature{$d_{gn}$}{Geometric mean size [\si{\meter}]}
\nomenclature{$\sigma_g$}{Geometric standard deviation [1]}

\nomenclature{$d_p$}{Particle diameter [m]}
\nomenclature{$\rho_p$}{Particle density [\si{\kilogram\ \meter^{-3}}]}

\nomenclature{$d_e$}{Needle diameter [m]}

\nomenclature{$k_b$}{Boltzmann constant [$\si{\joule\per\kelvin}$]}
\nomenclature{$Sc$}{Schmidt number, $Sc = \nu_a/D_B$ [1]}
\nomenclature{$Re$}{Reynolds number, $Re = U d_e/\nu_a$ [1]}
\nomenclature{$D_B$}{Brownian diffusion coefficient, $D_B = (C_C k_b T_a)/(3 \pi \mu_a d_p)$ [\si{\meter^2 \second^{-1}}]}
\nomenclature{$\lambda$}{Mean free path of the particle in the air, $\lambda$=$~0.066~\si{\micro \meter}$}
\nomenclature{$St$}{Stokes number, $St = \tau_p U / d_e$ [1]}
\nomenclature{$\tau_p$}{Particle relaxation time, $\tau_p = (\rho_p C_C d_p^2)/(18 \mu_a)$ [s]}
\nomenclature{$\tau_p^+$}{Dimensionless particle relaxation time, $\tau_p^+ = \tau_p u_f^2 /\nu_a$}

\printnomenclature[2cm]


\begin{thebibliography}{30}
\expandafter\ifx\csname natexlab\endcsname\relax\def\natexlab#1{#1}\fi
\expandafter\ifx\csname url\endcsname\relax
  \def\url#1{\texttt{#1}}\fi
\expandafter\ifx\csname urlprefix\endcsname\relax\def\urlprefix{URL }\fi

\bibitem[{Bae et~al.(2009)Bae, Jung, and Kim}]{BaeEtAl09}
Bae, S., Jung, C., Kim, Y., 2009. Development of an aerosol dynamics model for
  dry deposition process using the moment method. Aerosol Sci. Technol. 43~(6),
  570--580.

\bibitem[{Binkowski and Shankar(1995)}]{BinkowskiShankar95}
Binkowski, F., Shankar, U., 1995. The regional particulate matter model: 1.
  model description and preliminary results. J. Geophys. Res. 100~(D12),
  26191--26209.

\bibitem[{Greenshields(2015)}]{OpenFoamUserGuide}
Greenshields, C.~J., 2015. Open{FOAM} - {T}he {O}pen {S}ource {CFD} {T}oolbox -
  {U}ser's {G}uide. Version 3.0.0. CFD Direct Ltd.

\bibitem[{Gromke and Blocken(2015{\natexlab{a}})}]{GromkeBlocken15a}
Gromke, C., Blocken, B., 2015{\natexlab{a}}. Influence of avenue-trees on air
  quality at the urban neighborhood scale. {Part I}: Quality assurance studies
  and turbulent schmidt number analysis for {RANS} {CFD} simulations. Environ.
  Pollut. 196, 214--223.

\bibitem[{Gromke and Blocken(2015{\natexlab{b}})}]{GromkeBlocken15b}
Gromke, C., Blocken, B., 2015{\natexlab{b}}. Influence of avenue-trees on air
  quality at the urban neighborhood scale. {Part II}: Traffic pollutant
  concentrations at pedestrian level. Environ. Pollut. 196, 176--184.

\bibitem[{Hinds(1999)}]{Hinds99}
Hinds, W., 1999. Aerosol technology: Properties, Behavior, and Measurement of
  Airborne Particles, 2nd Edition. Wiley.

\bibitem[{Janhäll(2015)}]{Janhall15}
Janhäll, S., 2015. Review on urban vegetation and particle air pollution -
  deposition and dispersion. Atmos. Environ. 105, 130--137.

\bibitem[{Jeanjean et~al.(2015)Jeanjean, Hinchliffe, McMullan, Monks, and
  Leigh}]{JeanjeanEtAl15}
Jeanjean, A., Hinchliffe, G., McMullan, W., Monks, P., Leigh, R., 2015. A {CFD}
  study on the effectiveness of trees to disperse road traffic emissions at a
  city scale. Atmos. Environ. 120, 1--14.

\bibitem[{Jung et~al.(2003)Jung, Kim, and Lee}]{JungEtAl03}
Jung, C., Kim, Y., Lee, K., 2003. A moment model for simulating raindrop
  scavenging of aerosols. J. Aerosol Sci. 34~(9), 1217--1233.

\bibitem[{Katul et~al.(2004)Katul, Mahrt, Poggi, and Sanz}]{Katul04}
Katul, G., Mahrt, L., Poggi, D., Sanz, C., 2004. One- and two-equation models
  for canopy turbulence. Bound. Layer Meteor. 113, 81--109.

\bibitem[{Kenjereš and ter Kuile(2013)}]{KenjeresKuile13}
Kenjereš, S., ter Kuile, B., 2013. Modelling and simulations of turbulent
  flows in urban areas with vegetation. J. Wind Eng. Ind. Aerodyn. 123, 43--55.

\bibitem[{Koziol and Leighton(2007)}]{KoziolLeighton07}
Koziol, A., Leighton, H., 2007. The moments method for multi-modal
  multi-component aerosols as applied to the coagulation-type equation. Q. J.
  R. Meteorol. Soc. 133~(625), 1057--1070.

\bibitem[{Lalic and Mihailovic(2004)}]{LalicMihailovic04}
Lalic, B., Mihailovic, D., Apr. 2004. An empirical relation describing
  leaf-area density inside the forest for environmental modeling. J. Appl.
  Meteor. 43, 641--645.

\bibitem[{Litschke and Kuttler(2008)}]{LitschkeKuttler08}
Litschke, T., Kuttler, W., 2008. On the reduction of urban particle
  concentration by vegetation - a~review. Meteorol. Z. 17, 229--240.

\bibitem[{Mochida and Lun(2008)}]{MochidaLun08}
Mochida, A., Lun, I.~Y., 2008. Prediction of wind environment and thermal
  comfort at pedestrian level in urban area. J. Wind Eng. Ind. Aerodyn.
  96~(10–11), 1498--1527, 4th International Symposium on Computational Wind
  Engineering (CWE2006).

\bibitem[{Petroff et~al.(2008{\natexlab{a}})Petroff, Mailliat, Amielh, and
  Anselmet}]{PetroffEtAl08a}
Petroff, A., Mailliat, A., Amielh, M., Anselmet, F., 2008{\natexlab{a}}.
  Aerosol dry deposition on vegetative canopies. {Part I}: {Review} of present
  knowledge. Atmos. Environ. 42, 3625--3653.

\bibitem[{Petroff et~al.(2008{\natexlab{b}})Petroff, Mailliat, Amielh, and
  Anselmet}]{PetroffEtAl08b}
Petroff, A., Mailliat, A., Amielh, M., Anselmet, F., 2008{\natexlab{b}}.
  Aerosol dry deposition on vegetative canopies. {Part II}: A new modelling
  approach and applications. Atmos. Environ. 42~(16), 3654--3683.

\bibitem[{Petroff et~al.(2009)Petroff, Zhang, Pryor, and Belot}]{PetroffEtAl09}
Petroff, A., Zhang, L., Pryor, S., Belot, Y., 2009. An extended dry deposition
  model for aerosols onto broadleaf canopies. J. Aerosol Sci. 40~(3), 218--240.

\bibitem[{Pirjola et~al.(1999)Pirjola, Kulmala, Wilck, Bischoff, Stratmann, and
  Otto}]{PirjolaEtAl99}
Pirjola, L., Kulmala, M., Wilck, M., Bischoff, A., Stratmann, F., Otto, E.,
  1999. Formation of sulphuric acid aerosols and cloud condensation nuclei: an
  expression for significant nucleation and model comparison. J. Aerosol Sci.
  30~(8), 1079--1094.

\bibitem[{Raupach et~al.(2001{\natexlab{a}})Raupach, Briggs, Ford, Leys,
  Muschal, Cooper, and Edge}]{Raupach01a}
Raupach, M., Briggs, P., Ford, P., Leys, J., Muschal, M., Cooper, B., Edge, V.,
  2001{\natexlab{a}}. Endosulfan transport ii. modeling airborne dispersal and
  deposition by spray and vapor. J. Environ. Qual. 30~(3), 729--740.

\bibitem[{Raupach et~al.(2001{\natexlab{b}})Raupach, Woods, Dorr, Leys, and
  Cleugh}]{Raupach01}
Raupach, M., Woods, N., Dorr, G., Leys, J., Cleugh, H., 2001{\natexlab{b}}. The
  entrapment of particles by windbreaks. Atmos. Environ. 35, 3373--3383.

\bibitem[{Richards and Hoxey(1993)}]{RichardsHoxey93}
Richards, P., Hoxey, R., 1993. Appropriate boundary conditions for
  computational wind engineering models using the k-$\epsilon$ turbulence
  model. J. Wind Eng. Ind. Aerodyn. 46 \& 47, 145--153.

\bibitem[{Seinfeld and Pandis(2006)}]{SeinfeldPandis06}
Seinfeld, J., Pandis, S., 2006. Atmospheric Chemistry and Physics: From Air
  Pollution to Climate Change, 2nd Edition. A Wiley-Interscience publication.
  Wiley.

\bibitem[{Steffens et~al.(2012)Steffens, Wang, and Zhang}]{Steffens12}
Steffens, J., Wang, Y., Zhang, K., 2012. Exploration of effects of a vegetation
  barrier on particle size distributions in a near-road environment. Atmos.
  Environ. 50, 120--128.

\bibitem[{Tiwary et~al.(2005)Tiwary, Morvanb, and Colls}]{TiwaryEtAl05}
Tiwary, A., Morvanb, H., Colls, J., 2005. Modelling the size-dependent
  collection efficiency of hedgerows for ambient aerosols. J. Aerosol Sci. 37,
  990--1015.

\bibitem[{Tominaga and Stathopoulos(2007)}]{TominagaStathopoulos07}
Tominaga, Y., Stathopoulos, T., 2007. Turbulent {Schmidt} numbers for {CFD}
  analysis with various types of flowfield. Atmos. Environ. 41, 8091--8099.

\bibitem[{Vranckx et~al.(2015)Vranckx, Vos, Maiheu, and
  Janssen}]{VranckxEtAl15}
Vranckx, S., Vos, P., Maiheu, B., Janssen, S., 2015. Impact of trees on
  pollutant dispersion in street canyons: A numerical study of the annual
  average effects in {Antwerp}, {Belgium}. Sci. Total Environ. 532, 474--483.

\bibitem[{Whitby and McMurry(1997)}]{WhitbyMcMurry97}
Whitby, E., McMurry, P., 1997. Modal aerosol dynamics modeling. Aerosol Sci.
  Technol. 27~(6), 673--688.

\bibitem[{Whitby et~al.(1991)Whitby, McMurry, Shankar, and
  Binkowski}]{WhitbyEtAl91}
Whitby, E., McMurry, P., Shankar, U., Binkowski, F., 1991. Modal aerosol
  dynamics modeling. Tech. rep., U.S. Environmental Protection Agency, Research
  Triangle Park, NC 27711.

\bibitem[{Šíp and Beneš(2016)}]{SipBenes15b}
Šíp, V., Beneš, L., 2016. {CFD} optimization of a vegetation barrier. In:
  Karasözen, B., Manguoglu, M., Tezer-Sezgin, M., Göktepe, S., Ömür Ugur
  (Eds.), Numerical Mathematics and Advanced Applications - ENUMATH 2015.
  Springer International Publishing, Cham, to appear.

\end{thebibliography}
\end{document}